\documentclass[aps,floats,floatfix,amssymb,amsmath,prl,nofootinbib,twocolumn,superscriptaddress]{revtex4-1}

\usepackage{calc}
\usepackage{psfrag}
\usepackage{graphicx}
\usepackage{color}
\usepackage[dvipsnames]{xcolor}
\usepackage[unicode=true,pdfusetitle,
 bookmarks=true,bookmarksnumbered=false,bookmarksopen=false,
 breaklinks=false,pdfborder={0 0 0},backref=false,colorlinks=true]
 {hyperref}
\usepackage{geometry}
\usepackage{float}


\begin{document}

\title{Spin diffusion and spin conductivity in the 2d Hubbard model}

\author{Martin Ulaga}
\affiliation{Jozef Stefan Institute, Jamova 39, SI-1000, Ljubljana, Slovenia}
\author{Jernej Mravlje}
\affiliation{Jozef Stefan Institute, Jamova 39, SI-1000, Ljubljana, Slovenia}
\author{Jure Kokalj}
\affiliation{University of Ljubljana, Faculty of Civil and Geodetic
  Engineering,  Jamova 2, Ljubljana, Slovenia} 
\affiliation{Jozef Stefan Institute, Jamova 39, SI-1000, Ljubljana, Slovenia}

\begin{abstract}
  We study the spin diffusion and spin conductivity in the square
  lattice Hubbard model by using the finite-temperature Lanczos
  method. We show that the spin diffusion behaves differently from the
  charge diffusion and has a nonmonotonic $T$ dependence. This is due
  to a progressive liberation of charges that contribute to spin
  transport and enhance it beyond that active at low temperature due
  to the Heisenberg exchange. We further show that going away from
  half-filling and zero magnetization increases the spin diffusion,
  but that the increase is insufficient to reconcile the difference
  between the model calculations and the recent measurements on
  cold-atoms.
\end{abstract}
\maketitle

\section{Introduction}
\label{sec-main-i}
Non-saturating metallic resistivity $\rho$ that exceeds the
Mott-Ioffe-Regel (MIR) value (estimated with the scattering
length $l$ equal to the lattice spacing, $l \sim a$) is a
characteristic property of strongly correlated metals. Recent
theoretical work \cite{kokalj17,perepelitsky16} found that within
the Hubbard model, such behavior of conductivity $\sigma_c=1/\rho$
can be understood via Nerst-Einstein relation, $\sigma_c= D_c \chi_c$,
with a saturating diffusion constant $D_c$ and strongly suppressed
charge susceptibility at elevated temperatures.  Cold-atom
experiments \cite{brown19} have verified this behavior, which
established that at least the high-temperature regime
of the charge transport is fully understood.

Simultaneously, cold-atoms were also used to probe much less
explored quantities: the spin conductivity $\sigma_s$ and spin
diffusion constant $D_s$~\cite{nichols19} (the spin current that flows as a
response to the magnetic field or magnetization gradient). These
quantities are not only important from the theoretical point of view
(as the interplay between spins and charges lies at the heart of the
strong-correlation problem~\cite{anderson97,lee06}) but are important
also, e.g., for the applications in spintronics~\cite{zutic04,chumak15,hiroata20}, for
heat transfer\cite{zotos97,sologubenko01,hess01}, and further to
understand the behavior of the NMR relaxation
rate~\cite{sokol93,larionov04,yusuf07}. $\sigma_s$ or $D_s$ can be 
indirectly estimated via heat conductivity or NMR relaxation rate,
but also more directly through the spin injection technique \cite{johnson93,si97,
 ji04} and magnetization currents measurements \cite{meier03,
  cornelissen15}.

Importantly, in correlated materials, the spins and charges do not
behave alike and they exhibit a so-called spin-charge
separation~\cite{anderson97,lee06}, which means one cannot infer the
behavior of spin degrees of freedom (e.g., $\sigma_s$, $D_s$ and
$\chi_s$), based on the measurements of charge properties (e.g.,
$\sigma_c$, $D_c$ and $\chi_c$), and vice versa. The question that
arises is, how strong is the spin-charge separation in different
temperature ($T$) regimes?  Can one understand also the spin transport more
simply in terms of the Nernst-Einstein relation?  Does $D_s$ saturate and to what value? The cold-atom
experiment~\cite{nichols19} has not analyzed this behavior in detail,
and, intriguingly reported an inconsistency with the numerical
linked-cluster expansion (NLCE) method.

In this paper we consider these questions by solving the Hubbard model
with the finite-temperature Lanczos method (FTLM). We find that 
spin transport at high $T$ can be understood in terms of the
Nernst-Einstein relation, but the behavior is richer than the one for
the charge transport. $D_s$ has a nonmonotonic
$T$ dependence and experiences an increase at high
$T$ on the 
crossover between two saturated regimes: the lower-$T$ one at $T$
of the order of Heisenberg exchange $J$ with strong
spin-charge separation and the asymptotic high-$T$ one, where the spins
and charges behave alike. The scattering length remains of the order
of lattice spacing throughout this crossover and the behavior is
explained in terms of the evolution of effective velocity, instead.

Previously, the spin diffusion in 2d lattices was considered 
for the Heisenberg model, namely with high-$T$ frequency
moments \cite{bennett65,sokol93}, with theory of Blume and Hubbard
\cite{nagao98}, with interacting spin wave theory
\cite{sentef07,pires09} and with numerical~\cite{bonca95} and
Mori-Zwanzig~\cite{larionov04} approach to the $t$-$J$ model. The
spin diffusion in the Hubbard model was considered with Gaussian
extrapolation of short-time dynamics in the weak-coupling limit
\cite{kopietz98}, and with 
NLCE~\cite{nichols19}. To our knowledge, the $T$ dependence of $D_s$ in the Hubbard model, has not been
numerically calculated so far.

This paper is structured as following: we briefly present the method in
section \ref{sec-main-ii} then present the results for dynamical
conductivity in section \ref{sec-main-iii} and for spin diffusion
in section \ref{sec-main-iv}. We comment on a phenomenological
explanation for these results in section \ref{sec-main-v} and comment
on the effect of doping and magnetization in section
\ref{sec-main-vi}. Appendix \ref{sec-app-a} contains 
a derivation of the sum rule for dynamical spin conductivity
while Appendix \ref{sec-app-b}, \ref{sec-app-c} and \ref{sec-app-e} 
contain further details on the calculated quantities.

\section{Model and method}
\label{sec-main-ii}
We consider the Hubbard  model on the square lattice,
\begin{equation}
 H = -t\sum_{\langle i,j\rangle,s} c^\dagger_{i,s} c_{j,s}+U\sum_{i}
 n_{i,\uparrow}n_{i,\downarrow},
\label{eq_ham}
\end{equation}
where $c^\dagger_{i,s}/c_{i,s}$ create/annihilate an electron of spin
$s$ (either $\uparrow$ or $\downarrow$) at the lattice site $i$. $t$
is the hopping amplitude between the nearest neighbors and we use it
as the energy units. We further set $\hbar=k_B=e=g\mu_B=1$. We denote
the lattice 
constant with $a$.

We solve the model with FTLM
\cite{jaklic00,prelovsek13,kokalj13}, which uses the Lanczos algorithm
to obtain approximate eigenstates of the Hamiltonian and additional
sampling over the initial random vectors to treat finite-$T$
properties on small clusters. We use $N=4\times 4$ cluster. 
To reduce the finite-size effects that appear at $T\lesssim t$
we further employ averaging over twisted boundary conditions and use
the grand canonical ensemble. This allows us to reach reliable
results, e.g, for $U=10t$, at $T\gtrsim 0.2t$ for the thermodynamic
quantities (e.g., spin susceptibility $\chi_s$), and at $T\gtrsim
0.8t$ for dynamical quantities (e.g., dc spin conductivity
$\sigma_s$). We omit low-$T$ results for which, due to finite size,
dynamical spin stiffness exceeds 1\% of the total spectral weight. 

To calculate $D_s$ (and analogously the charge
diffusion constant $D_c$) we use the Nernst-Einstein relation (see,
e.g., Refs.~\onlinecite{kopietz98,pakhira15} for derivation) 
\begin{equation}
\sigma_s=D_s \chi_s . 
\end{equation}
The dc spin conductivity $\sigma_s$ is calculated as the $\omega=0$
value of the dynamical spin conductivity $\sigma_s(\omega)$, which is
directly evaluated as the current-current correlation function\cite{jaklic00, kokalj17}.
$\sigma_s(\omega)$ for finite cluster consists of delta functions
which need to be broadened. This introduces some uncertainty in
dynamical results and we estimate the uncertainty due to finite-size
effects and broadening to be $\lesssim 10\%$.

\begin{figure}[ht!]
 \begin{center}
   \includegraphics[ width=0.99\columnwidth]{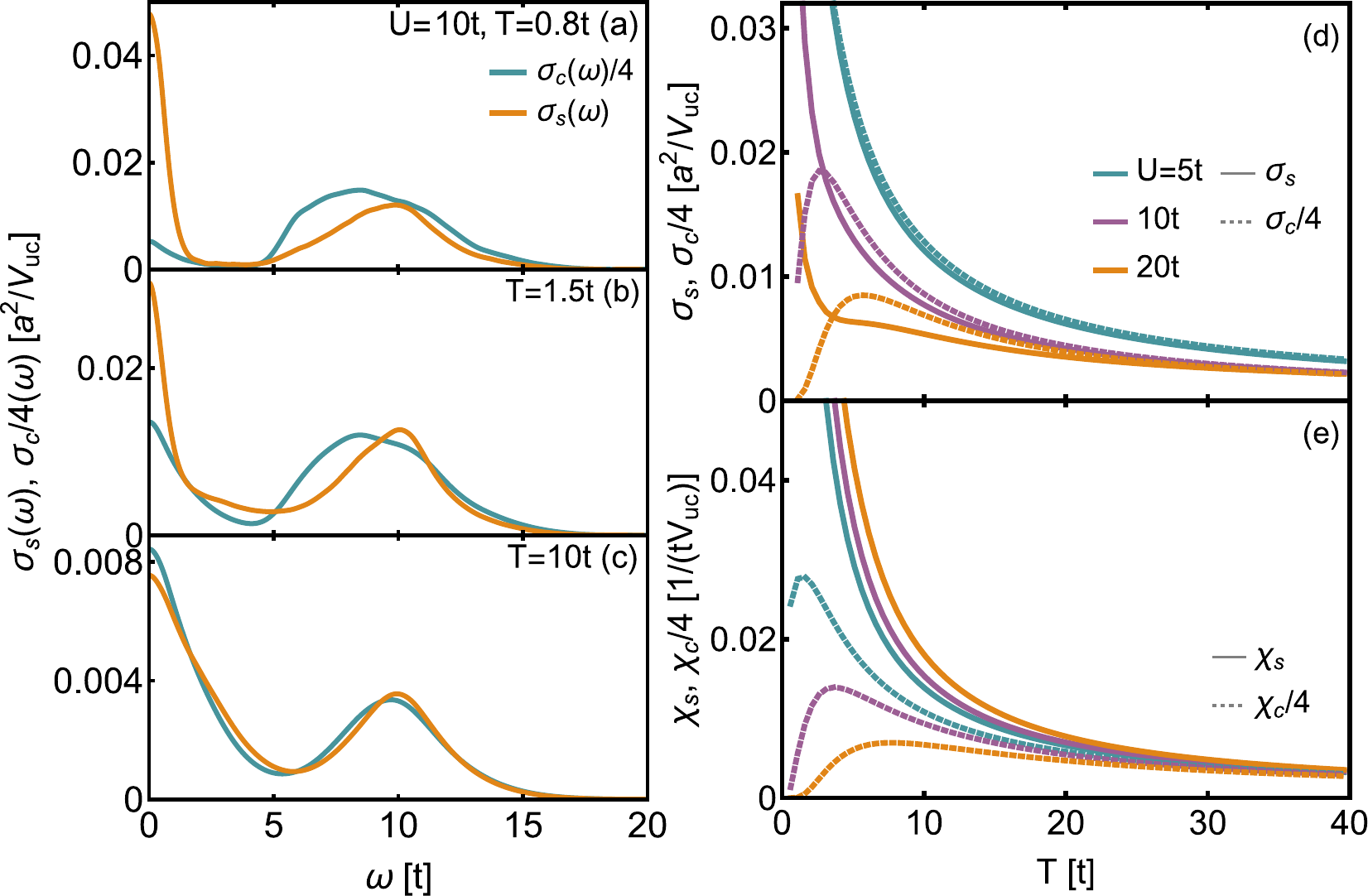} 
 \end{center}
 \caption{Dynamical spin conductivity $\sigma_s(\omega)$ and charge
   conductivity $\sigma_c(\omega)$. At low $T$ (a) $\sigma_s(\omega)$
   shows a large Drude-like peak and the conducting character, while
   $\sigma_c(\omega)$ shows an insulating behavior with only a small value
   at $\omega\sim 0$.  At intermediate temperature $J<T<U$ (b)
   low-$\omega$ peak in $\sigma_s(\omega)$ develops a shoulder like structure.  At
   high $T$ (c) $\sigma_s(\omega)$ and $\sigma_c(\omega)$ become
   similar.  Comparison (d)
   of $\sigma_s$ and $\sigma_c$, which are similar for $T\gtrsim U/2$,
   and comparison (e) of $\chi_s$ and $\chi_c$.  Results are for
   half-filling ($p=1-n=0$).
  }
 \label{fig1a}
\end{figure}

\section{Dynamical conductivity}
\label{sec-main-iii}
In Fig.~\ref{fig1a}(a-c) we show the
dynamical spin $\sigma_s(\omega)$ and charge $\sigma_c(\omega)$ conductivities
for the half-filled Hubbard  
model at half-filling for $U=10t$. The two quantities behave very
similarly at high $T$, and both display a low-$\omega$
``Drude'' peak, and a high-$\omega$ peak at $\omega\sim U$ due to
transitions to Hubbard
bands.

On cooling down, there is a growing degree of the spin-charge
separation. The charge transport is depleted and the low-$\omega$
peak in $\sigma_c(\omega)$ is suppressed, corresponding to the Mott insulating
regime. Conversely, in the spin conductivity, a peak develops at
$\omega \sim 0$, corresponding to a spin-metallic regime in the spin
sector (see also Fig.~\ref{fig1a}(d) for the $T-$dependence of the dc
transport).  The two quantities $\sigma_s(\omega)$ and $\sigma_s(\omega)$
are actually never fully independent
as they are related by the f-sum rule. Namely, their integrals over
frequency are equal up to a factor of 4~\cite{fishman02} as we show in Appendix \ref{sec-app-a}.
This also
means that at low $T$, parts of the spectral weight in
$\sigma_s(\omega)$ is in comparison to $\sigma_c(\omega)$ removed from
the Hubbard band to accommodate the increase at low $\omega$.  One
can relate the behavior of $\sigma_{c,s}(\omega)$ to charge and spin
fluctuations, e.g., $\sigma_{c,s}(\omega)=\lim_{q\to 0}
\frac{\omega}{q^2}\textrm{Im}[\chi_{c,s}(q,\omega)]$. Here,
$\chi_{c,s}(q,\omega)$ are the dynamical susceptibilities. Relative to
the charge sector, on lowering $\omega$ the spin fluctuations are
first suppressed and then increased.  One can also notice, that
$\sigma_s(\omega)$ at intermediate $T$ develops an interesting 
two peak like structure at $\omega\sim 0$ (see
Fig.~\ref{fig1a}(b)), i.e., a sharper peak on top of the broader peak
both centered at $\omega=0$. Notice also that whereas the key
distinction between the behavior of spins and charges could be
expected at the energy scales of the order Heisenberg exchange
$J=4t^2/U$, the two conductivities differ also at larger energy
scales.

It is worth mentioning, that the difference between 
$\sigma_s(\omega)$ and $\sigma_c(\omega)$ vanishes at the bubble level
and appears due to the vertex correction (see, e.g.,
Refs.~\onlinecite{fishman02, vucicevic19, vranic20}). Their
difference is therefore a direct indication of the importance of
vertex corrections.

In Fig.~\ref{fig1a}(e) we show also the $T$
dependence of spin and charge susceptibilities $\chi_{s,c}$.
One sees a clear distinction between $\chi_c$, that decreases at
low-$T$ indicating a charge gap and large values of $\chi_s$
indicating large local-moment and Curie-Weiss-like behavior. See
Appendix \ref{sec-app-c} for more details. 

\begin{figure}[ht!]
 \begin{center}
   \includegraphics[ width=0.48\textwidth]{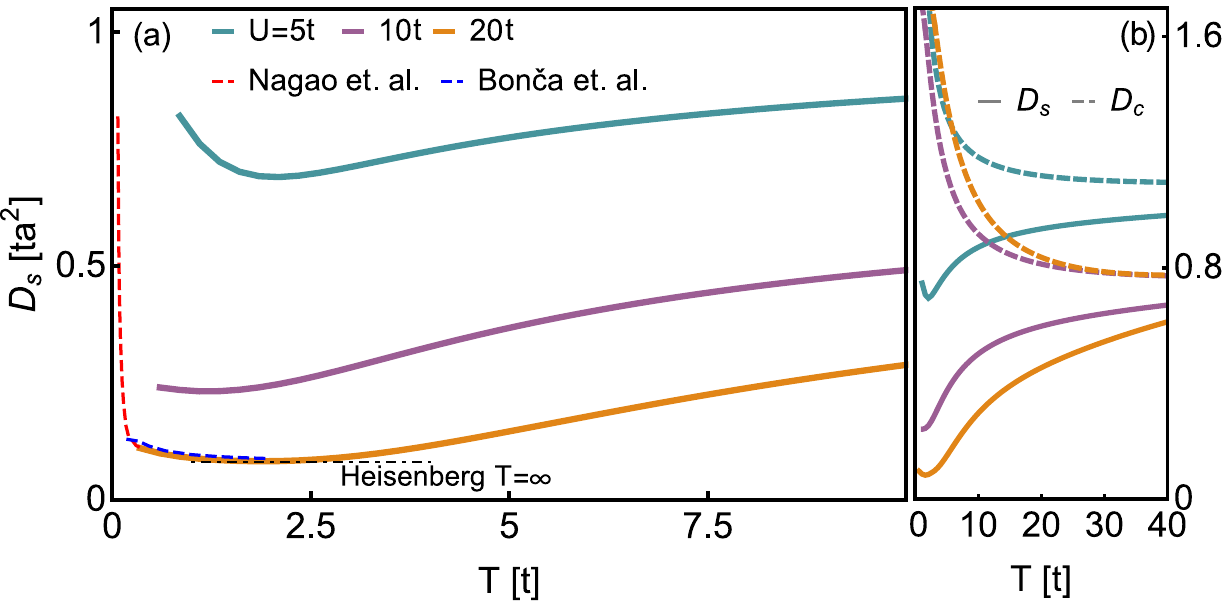} 
 \end{center}
 \caption{
   (a) Spin diffusion constant $D_s$ for half-filled Hubbard
model for $U/t=5,10,20$.  
$D_s$ decreases with increasing $U$ and shows a nonmonotonic $T$
dependence. As $T\to 0$ one expects diverging $D_s$ due to increased
spin coherence\cite{bonca95,nagao98,kopietz98}, which shows up as a
small upturn in our results. The divergence is indicated by showing the result
of a Blume and Hubbard theory for the Heisenberg model from Nagao et
al.~\cite{nagao98} for $U=20t$. 
 In the intermediate regime $J\ll T\ll U$, $D_s$ is
minimal and for large $U$ shows a saturating behavior corresponding to the
high-$T$ Heisenberg result~\cite{sokol93}  and agrees with FTLM result
for the Heisenberg model (Bon\v ca et al.~\cite{bonca95}).
(b) As $T$ increases
further, $D_s$ shows an increase and saturates for $T\to \infty$ at the
same value as the charge diffusion $D_c$ and in accordance with the
standard Mott-Ioffe-Regel limit.
}
\label{fig2}
\end{figure}

\section{Spin diffusion}
\label{sec-main-iv}
We now turn to the $T$ dependence of
the diffusion constant.  Fig.~\ref{fig2}(a) shows $D_s$ vs. $T$
for several $U$. Even though the spin conductivities
are all metallic, $D_s$ shows an unusual nonmonotonic $T$
dependence. Unlike in the case of charge transport in metals (where
$D_c$ monotonously increases with lowering $T$), $D_s$ initially drops,
reaches a minimum and only at lowest available $T$ starts to grow. The
growth of $D_s$ in this low-$T$ regime can be 
discussed in terms of a growing correlation length and associated
coherence of spin-waves in the Heisenberg model\cite{nagao98,
  chakravarty89} 
and associated longer mean free path
$l_s$ (as $D_s\sim v_s l_s/2$ with $v_s$ a characteristic spin
velocity). To indicate the expected behavior, we supplement our
results in Fig.~\ref{fig2}(a)) with a result of Nagao et
al.\cite{nagao98} for the Heisenberg model.

At intermediate $T$ ($\sim 2t$), $D_s$ reaches a minimum with
indications for intermediate saturating behavior seen for larger
$U$. In the regime of $T\ll U$ for large $U$, the behavior becomes
that of the Heisenberg model and is therefore entirely
controlled by $J$.  The calculated $D_s$ hence agrees with the results
from Heisenberg model, including high-$T$
moments expansion \cite{sokol93}, numerical FTLM \cite{bonca95} and
with self-consistent Blume and Hubbard theory \cite{nagao98}. $D_s$
further shows in a regime $J \ll T\ll U$ a saturation towards the
high-$T$ limit of the Heisenberg model~\cite{sokol93,bonca95} with
$D_s\sim 0.40Ja^2$, see results for $U=20t$ in Fig.~\ref{fig2}(a). Such
high-$T$ value can be understood in terms of a ``spin Mott-Ioffe-Regel''
value by approximating the spin-wave velocity to
$v_s \sim \frac{4Ja}{\sqrt 2 \pi}$ and $l_s$ to
minimal or MIR limiting value $l_s\sim a$. This leads to
$D_s\sim v_s l_s/2 \sim 0.45 J a^2$.

With further increase of $T$ towards $U$, $D_s$ remarkably
increases. This is in contradiction with a naive 
expectation of decreasing mean free path (increasing scattering rate)
and therefore decreasing $D_s$ with increasing $T$. The reason for
this can be found in the increase of empty and doubly occupied sites
allowing for new conducting and diffusive mechanism of spin in terms
of electron hopping, in addition to exchange mechanism dominating
at low $T$.

Fig.~\ref{fig2}(b) additionally shows the charge diffusion
 constants compared to $D_s$ in a broader $T$ range. One
 notices that $D_c$ decreases on heating up
 (which is the standard behavior) and that $D_c$ and $D_s$ approach
 each other at very high $T$. There, both $D_s$ and $D_c$ saturate at the
 usual MIR limit $D\sim at^2$.

 It is interesting to observe that in a broad temperature regime
 $T\gtrsim U/2$ the conductivities
 $\sigma_s$ and $\sigma_c$ differ little ($<20$\%) (Fig.~\ref{fig1a}(d)),
 while the corresponding diffusion constants differ by a factor of
 almost 2, (Fig.~\ref{fig2}(b)). The difference between $D_s$ and $D_c$ is
 compensated by the inverse difference between susceptibilities $\chi_s$
 and $\chi_c$ (Fig.~\ref{fig1a}(e)) to give similar conductivities in the whole $T\gtrsim
 U/2$ regime. This suggest an intriguing relationship between the
 diffusion (dynamic) and susceptibility (static property).

In passing we mention also  that in the high-$T$ limit ($T\gg U$), $D_s$
 does not show the $t^2/U$ scaling, as suggested from the moment
 expansion analysis \cite{kopietz98}. This can be traced back to the
 two peak structure in the dynamical spin conductivity
 (Fig.~\ref{fig1a}) with the upper Hubbard peak positioned at
 $\omega \sim U$, which is quite challenging to correctly reproduce
 from the frequency moments.

 \section{Mean free path}
 \label{sec-main-v}
\begin{figure}[ht]
\begin{center}
\includegraphics[width=0.95\columnwidth]{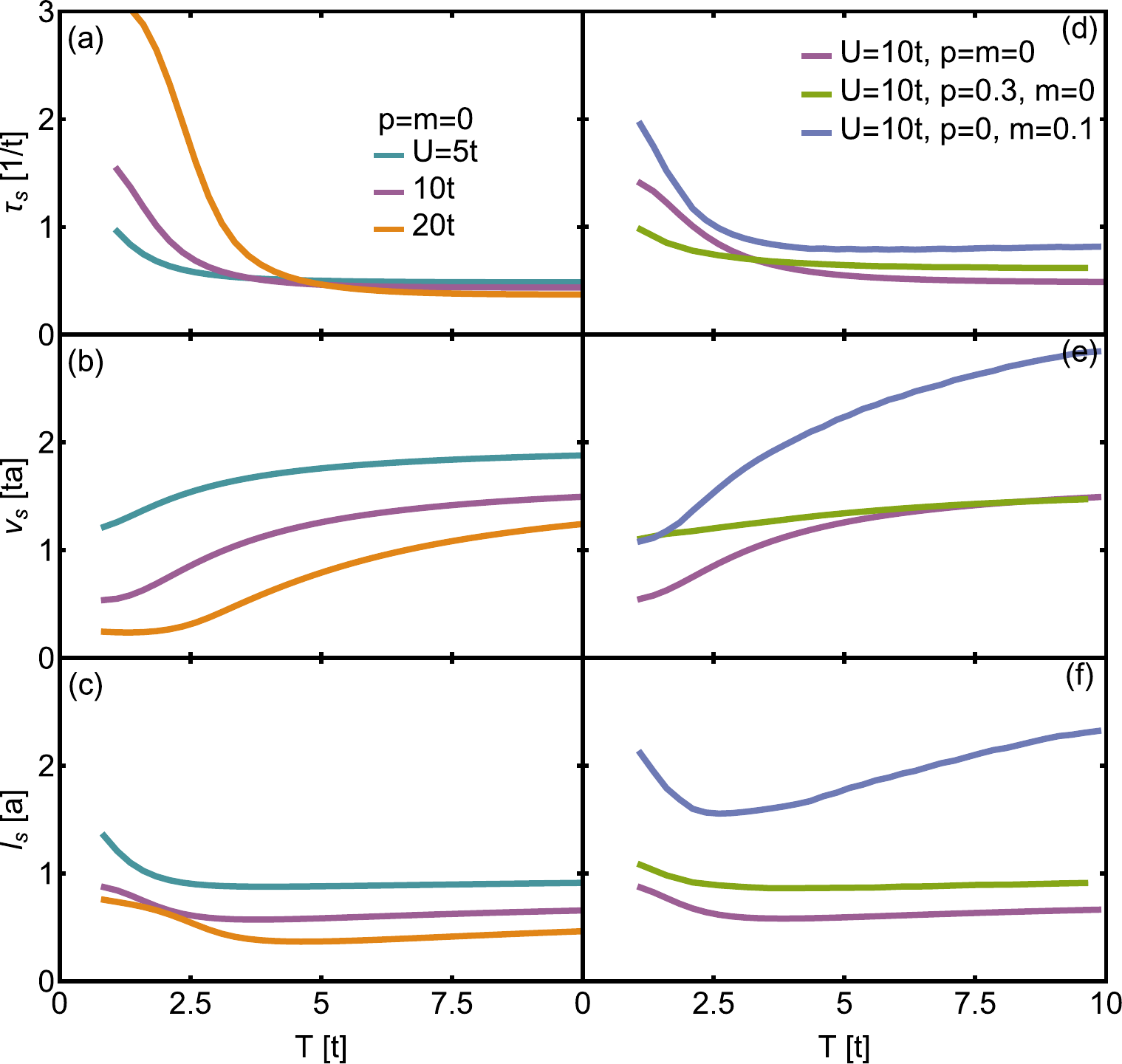}
\caption{$T$ dependence of spin scattering time
  $\tau_s$, spin velocity $v_s$ and spin mean free path $l_s$ for several
  choices of interaction strength $U$, hole doping $p$ and
  magnetization $m$.}
\label{fig3}
\end{center}
\end{figure}

It is instructive to investigate the phenomenology of our results in
more details. From the width of the low frequency peak in
$\sigma_s(\omega\sim 0)$ (e.g., Fig.~\ref{fig1a}(a-c)) we estimate 
a spin scattering time $\tau_s$ (see also Appendix \ref{sec-app-e}). Then we use
a simple relation
$D_s\sim \frac{v_s^2\tau_s}{2}$ and the values of $D_s$  to estimate the spin velocity
$v_s$ and further the spin mean free path $l_s\sim v_s\tau_s$.  The results are plotted on
Fig.~\ref{fig3}, and reveal evolution between two regimes, a
lower-$T$ one governed by the scale $J$ and a higher-$T$ one governed
by $t$. 
 $\tau_s$ is seen to exhibit a pronounced increase below $T\sim U/2$,
which coincides with a sharp structure of width $1/\tau_s\sim J$
emerging in the dynamical spectra. At $T\gtrsim U/2$ it saturates at
the value of order $1/\tau_s\sim t$. 
The extracted characteristic velocity starts at values $v_s \sim Ja$ at
low $T$ and increases monotonically to the value of   $v_s\sim
ta$. The low-$T$ estimate of $v_s\sim 0.6 ta$ for $U=10t$ 
(Fig.~\ref{fig3}(b)) is remarkably close to the estimate of
$v_s\sim1.6Ja\sim 0.64ta$ within the Heisenberg model\cite{kim98}, in
particular, since we used a rough approximation $D_s\sim v_s^2 \tau_s
/2$. 

Conversely, $l_s$ is to a good approximation $T$ independent and close
to the lattice spacing. It shows only a moderate increase at lowest
$T$. The spin transport is thus characterized by a saturated
scattering length throughout the considered $T$ regime (except at
lowest $T$) and the effects seen in the diffusion constant are
explained in terms of progressive unbinding of the charge degrees of
freedom that progressively increase the corresponding velocity to a
value given by $t$ instead of $J$. In the half-filled case, this
increase of $v_s$ is the main reason for the increase of
$D_s\sim v_s l_s/2$ with increasing $T$ (see Fig.~\ref{fig2}). We note
that the dependence of $l_s$ on $U$ in Fig.~\ref{fig2}(c) is smaller
at higher $T$($>U$) and $l_s$ is closer to $a$ for all $U$.

\begin{figure}[ht!]
 \begin{center}
   \includegraphics[ width=1\columnwidth]{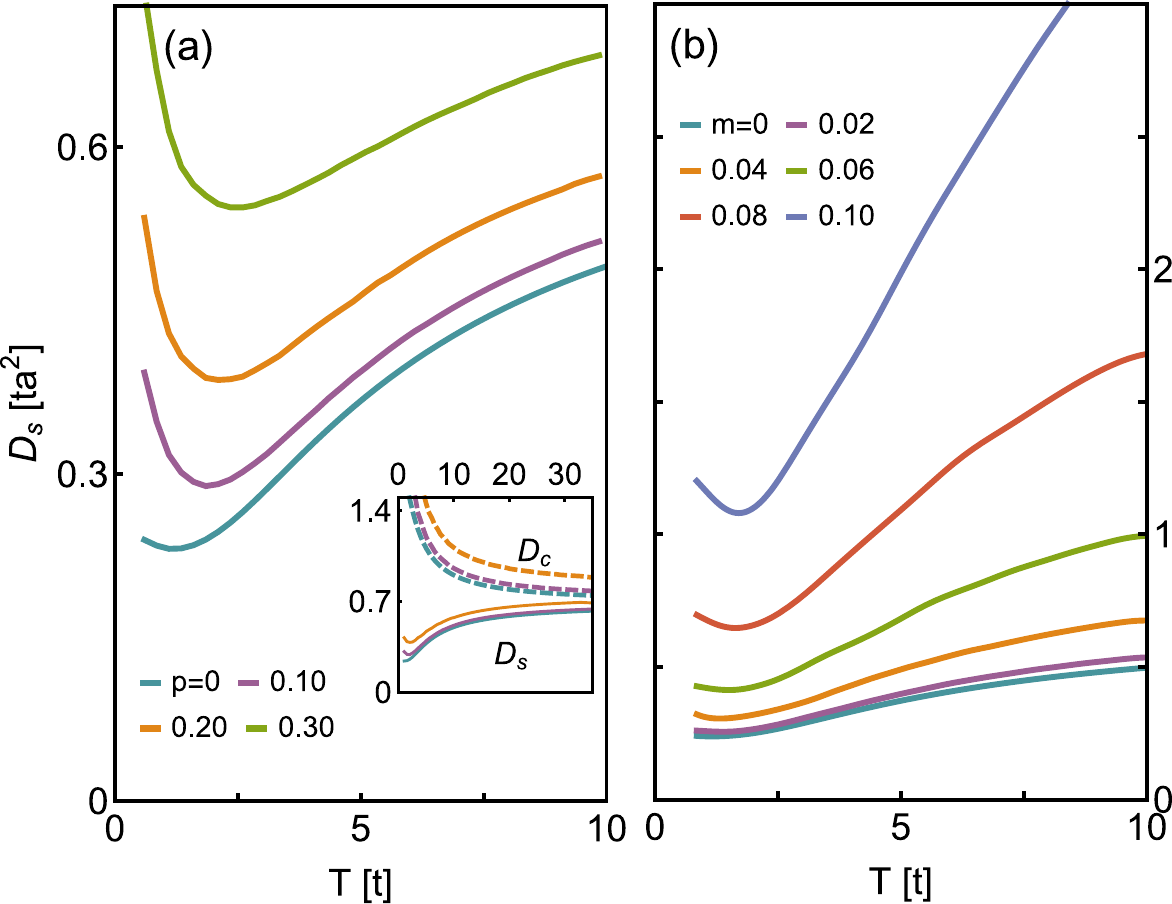} 
 \end{center}
 \caption{(a) Increase of spin diffusion $D_s$ with doping $p$.
 (Inset) In the whole shown $T$ and $p$ regime $D_s$ is smaller and behaves qualitatively
   differently than $D_c$, even at largest $p=0.3$. They become
   similar only in the very high  $T$ limit. $U=10t$.
   (b)  $D_s$ vs. $T$ for several values of magnetizations
   $m$. With increasing $m$, $D_s$ increases in whole $T$ range.
 $U=10t$, $p=0$.}
 \label{fig4}
\end{figure}

\section{Doping and magnetization effect}
\label{sec-main-vi}
How is this picture modified
at finite dopings and magnetizations?  Because moving electrons carry
both spin and charge, one could expect that away from half-filling,
when the system is metallic, $D_s$ and $D_c$ behave more similarly. In
Fig.~\ref{fig4}(a) we show the behavior of $D_s$ for several dopings
$p=1-n$. $n$ is the electron density. One clearly observes the
increase of $D_s$ with increasing $p$, which is understood as opening
of a new conducting channel via hopping of itinerant electrons or
holes. The increase is particularly strong at lowest calculated $T$
and the indication of diverging $D_s(T\to 0)$ becomes more
apparent. In doped case, $D_s$ approaches $D_c$, but $D_s$ and $D_c$
still behave distinctly (Fig.~\ref{fig4} (inset)), with $D_s$ having
much smaller values and a pronounced minimum at intermediate $T$.
Thus, doping diminishes the degree of spin-charge separation, but does
not wash it out completely.  At low-$T$, $D_s$ is much smaller than
$D_c$ and the spin transport is less coherent than the charge
transport, possibly due to stronger coupling to low lying spin excitations.
The extracted $l_s$
remains roughly $T$ independent (Fig.~\ref{fig3}(f)) and is
somewhat larger, but still $l_s\sim a$. The extracted $v_s$ and $\tau_s$ show less $T$ dependence than
in the undoped case (Fig.~\ref{fig3}(d-e)). 

The dependence on magnetization $m=\langle S_{z,tot}\rangle/N$ is
shown on Fig.~\ref{fig4}(b). It is found to be initially weak but
becomes strong with increasing $m$.  The results for $m=0.1$ deviate
significantly from nonmagnetized ones, e.g., $D_s$ is increased by
more than a factor of 4.  The underlying physics differs from the case
of charge doping: with increasing $m$ one stays in, or even goes
deeper into, the Mott insulating phase (see Fig.~1 and 2 in
Ref.~\onlinecite{prelovsek15}).  The increase of $D_s$ is therefore
not due to mobile electron-like particles, but rather due to weaker
scattering of spin waves and their longer $l_s$. This is indeed
revealed in Fig.~\ref{fig3}(f), where $l_s$ is increased due to
increase of both $v_s$ and $\tau_s$.  With increasing $m$ the Hubbard
interaction becomes less effective and one approaches the limit of
noninteracting spins or single noninteracting holon-doublon
pair\cite{prelovsek15} (the limit of only one $\downarrow$ spin in the
background of $\uparrow$ spins).  The strong increase of $D_s$ with
increasing $T$ at $T \gtrsim 2t$ is mainly due to increasing $v_s$,
but surprisingly, also $l_s$ moderately increases with increasing $T$
as well.

\section{Discussion}
\label{sec-main-vii}
\begin{figure}[h]
\begin{center}
\includegraphics[width=0.28\textwidth]{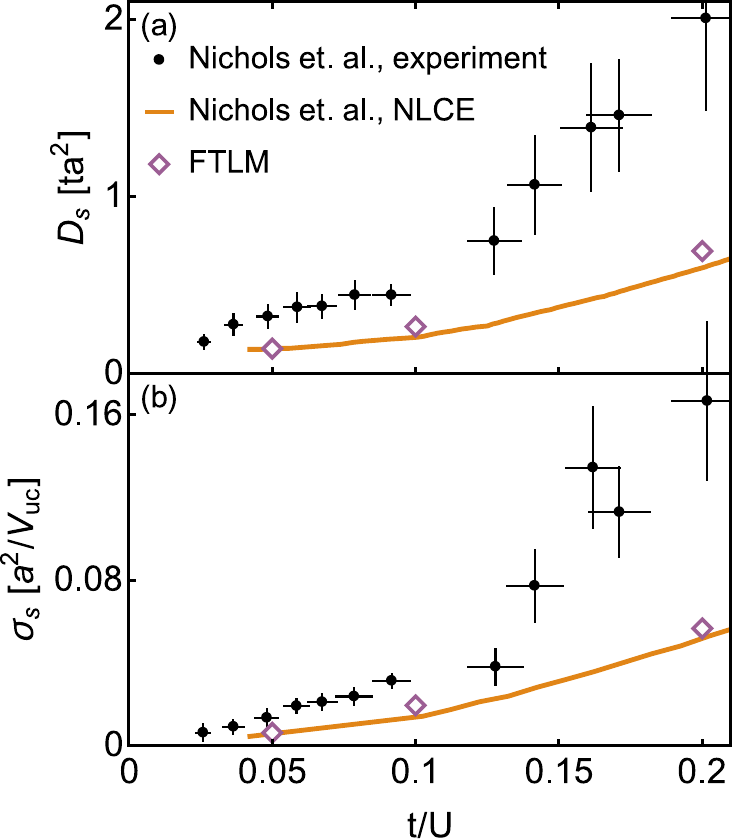}
\caption{Comparison of our FTLM results with experimental data\cite{nichols19} and
  numerical NLCE results\cite{nichols19}. FTLM results are for fixed
  entropy density of $1.1k_B$, which was estimated for 
  experiment  and taken also in the NLCE calculations.}
\label{fig:comparison_exp}
\end{center}
\end{figure}

  We compared our
  results to the $D_s$ and $\sigma_s$ measured in the cold atom
  experiment\cite{nichols19}, see Fig.~\ref{fig:comparison_exp}, 
  where the measured $D_s$ and $\sigma_s$ are found to be by about 
  a factor  of 2 higher than the   NLCE results for the Hubbard model.
  Our results agree well with the NLCE  results.  As discussed in Sec.~VI.,
  $D_s$ and $\sigma_s$ are increased by magnetization. We estimate that
  the  experimental deviation from half-filling and zero magnetization
  are too small to understand the discrepancy between the
  numerics and experiment in those terms.
  
  We also compare  our results (see Appendix \ref{sec-app-d}) with the diffusion bound $D_s>v_s^2/T$ suggested from holographic
  duality\cite{hartnoll15}. Using our rough estimate for $v_s$ extracted from 
  $D_s$ and $\tau_s$, we observe that the bound is obeyed in most of the 
    explored parameter regime except at small $U$, where it is mildly
    violated. We also observe that $D_s$ does not follow or has a $T$ dependence close to the proposed bound in the calculated $T$ regime. See also Ref.~\onlinecite{pakhira15}.

\section{Conclusions}
\label{sec-main-viii}
We have shown that $D_s$ has a striking
nonmonotonic $T$-dependence, 
which can be understood in terms
of a crossover from a low-$T$ spin-wave dominated regime to a high-$T$
asymptotic regime via an intermediate strong spin-charge separation
regime, where it nearly 
saturates according to the high-$T$ Heisenberg model result at  $D_s
\sim Ja^2$. In both of these high-$T$ regimes, the 
scattering length is of the order lattice spacing and the conductivity
is strongly reduced due to decreased $\chi_s$. In analogy to the case of
charge transport, these regime can be referred to as a {\it ``spin
  bad metal''}.  The action in 
$D_s$ is governed by the spin velocity that evolves from $J$ to $t$
up to the asymptotic high-$T$ regime, where
spins and charges behave alike and $D_s \sim D_c \sim t a^2$.

In the whole regime of $T\lesssim U/2$ (with or without doping or
magnetization) the spins and charges behave differently. An interesting
open question is better characterization of the behavior at lower $T$.
Away from half-filling in the well-defined quasiparticle
regime one expects similar behavior of spin and charge transport.
Further studies in real materials, e.g., with techniques like spin injection
\cite{johnson93,si97, ji04} or magnetization currents measurements
\cite{meier03, cornelissen15} would be highly valuable. Better
understanding of the  spin transport would also shed light on thermal
conductivity and NMR relaxation rate\cite{takigawa96}, for which our
results suggest a nonmonotonic-in-$T$ diffusive contribution. 

\section*{Acknowledgments}
We acknowledge helpful discussion with Peter Prelov\v{s}ek, Friedrich Krien,
Matthew A. Nichols and Martin W. Zwierlien. This work was supported by the
Slovenian Research Agency (ARRS) under Program No. P1-0044.

 \appendix{

\section{Sum rule}
\label{sec-app-a}
To derive the spectral sum rule for the dynamical spin conductivity we
employ the approach by B. S. Shastry\cite{shastry09}.
The sum rule is given by the expectation value of the stress tensor $\mathcal{T}$
\begin{equation}
\int_{-\infty}^\infty \sigma_{c,s}(\omega)d\omega=\frac{\pi \langle \mathcal{T}_{c,s} \rangle} {N}.
\end{equation}
Supscripts $c$ and $s$ stand for charge and spin quantities,
respectively. 
The charge stress tensor is given by\cite{shastry09}
\begin{equation}
\mathcal{T}_c=-\lim_{q\to 0}\frac{d}{dq} [j_c(q),n(-q)],
\label{eq_tauc}
\end{equation}
where the charge current and charge density operators are
\begin{eqnarray}
j_c(q)&=& i t \sum_{i,\delta,s}\delta_x c^\dagger_{i+\delta,s}c_{i,s}
          \textrm{e}^{iq(R_{i,x}+\delta_x/2)}, \label{eq_jc}\\
n(q)&=&\sum_{i,s} n_{i,s} \textrm{e}^{iq R_{i,x}}.\label{eq_n}
\end{eqnarray}
Here, the $x$ direction of current and of the wave vector $q$ is explicitly
used. $\delta$ goes over all nearest neighbors and $\delta_x$ denotes
the spacial distance to neighboring site in $x$ direction. $R_{i,x}$
is the $x$ coordinate of site $i$. 
Similarly one can write the spin stress tensor
\begin{equation}
\mathcal{T}_s=-\lim_{q\to 0}\frac{d}{dq} [j_s(q),m(-q)],
\label{eq_taus}
\end{equation}
and the spin current and magnetization density operators,
\begin{eqnarray}
j_s(q)&=& i t \sum_{i,\delta,s} s \delta_x c^\dagger_{i+\delta,s}c_{i,s}
          \textrm{e}^{iq(R_{i,x}+\delta_x/2)}, \label{eq_js}\\
m(q)&=&\sum_{i,s} s n_{i,s} \textrm{e}^{iq R_{i,x}}. \label{eq_m}
\end{eqnarray}
Here the factor $s$ in the sums for $j_s(q)$ and $m(q)$ is taken to be
$1/2$ for $\uparrow$ spins and $-1/2$ for $\downarrow$ spins. 
Evaluating the commutation in expressions for $\mathcal{T}_c$ in
Eq.~\ref{eq_tauc} and for $\mathcal{T}_s$ in Eq.~\ref{eq_taus} together with
$q$-derivative and limit, one obtains similar expressions for
$\mathcal{T}_c$ and $\mathcal{T}_s$. 
\begin{eqnarray}
\mathcal{T}_c&=&t\sum_{i,\delta,s} \delta_x^2 c_{i+\delta,s}^\dagger c_{i,s},\label{eq_tauc1}\\
\mathcal{T}_s&=&t\sum_{i,\delta,s} s^2 \delta_x^2 c_{i+\delta,s}^\dagger
          c_{i,s}.
\label{eq_taus1}
\end{eqnarray}
The reason for the only difference of factor $s^2=1/4$ in expression
for $\mathcal{T}_s$ (\ref{eq_taus1}), in comparison to $\mathcal{T}_c$
(\ref{eq_tauc1}), is the commutation of spin $\uparrow$ and
$\downarrow$ operators and no spin mixed terms in the sums for
currents and density operators
(Eqs. \ref{eq_jc}, \ref{eq_n}, \ref{eq_js} and \ref{eq_m}). The
expectation values of $\mathcal{T}_c$ and $\mathcal{T}_s$ correspond to the
expectations values of the kinetic energy for our nearest-neighbor
Hubbard model \cite{jaklic00}
\begin{eqnarray}
\langle \mathcal{T}_c \rangle &=& \frac{a^2}{2} \langle H_{kin}\rangle,\\
\langle \mathcal{T}_s \rangle &=& \frac{a^2}{8} \langle H_{kin}\rangle.
\end{eqnarray}
This leads to the (optical) sum rule given in the main text.
\begin{eqnarray}
\int_{-\infty}^\infty \sigma_c(\omega)d\omega&=& \frac{\pi a^2}{2N}\langle
  H_{kin} \rangle,\\
\int_{-\infty}^\infty \sigma_s(\omega)d\omega&=& \frac{\pi a^2}{8N}\langle
  H_{kin} \rangle,
\end{eqnarray}
with the difference only of a factor of 1/4. 

\section{Finite size effects}
\label{sec-app-b}
To estimate the error due to finite size effects, we plot the optical 
spin conductivity for different system sizes on Fig.~\ref{fig:sup_size}. We find
that they differ substantially only around $\omega=0$. This is the basis for the
error estimate given in the main text.

\begin{figure}[h]
\begin{center}
\includegraphics[width=0.4\textwidth]{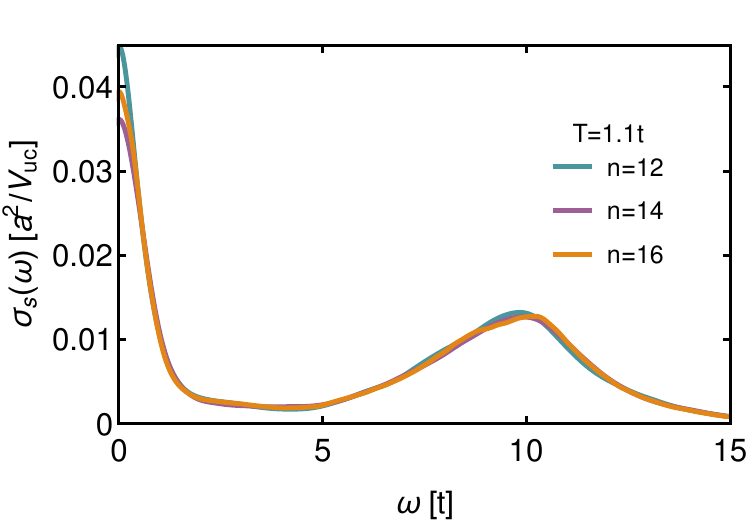}

\caption{Comparison of the optical spectrum of spin conductivity 
obtained on different size clusters. We use 16 different periodic boundary
conditions for $N=12$ and $N=14$ and 8 for $N=16$.}
\label{fig:sup_size}
\end{center}
\end{figure}

\section{Resistivity}
\label{sec-app-c}
It is instructive to inspect also the inverse conductivity, namely
the resistivity $\rho=1/\sigma$. See Fig.~\ref{fig:sup_rho}(a). Linear in
$T$ resistivity is easily recognized for both $\rho_c$ and $\rho_s$ at
high $T$ and $\rho_c$ shows a pronounced crossover into the insulating
regime at low $T$, while $\rho_s$ remains metallic. Spin susceptibility $\chi_s$
shows (Fig.~\ref{fig:sup_rho}(b)) an increase below $T\sim U$ and approaches 
the value of the high-$T$ Heisenberg limit with
$T\chi_s=1/4$, which is expected for large $U$ and $J\ll T\ll
U$. In the ultra high $T$ limit, $T\gg U$, both $T\chi_s$ and
$T\chi_c/4$ approach the atomic limit of $1/8$.

\begin{figure}[h]
\begin{center}
\includegraphics[width=0.4\textwidth]{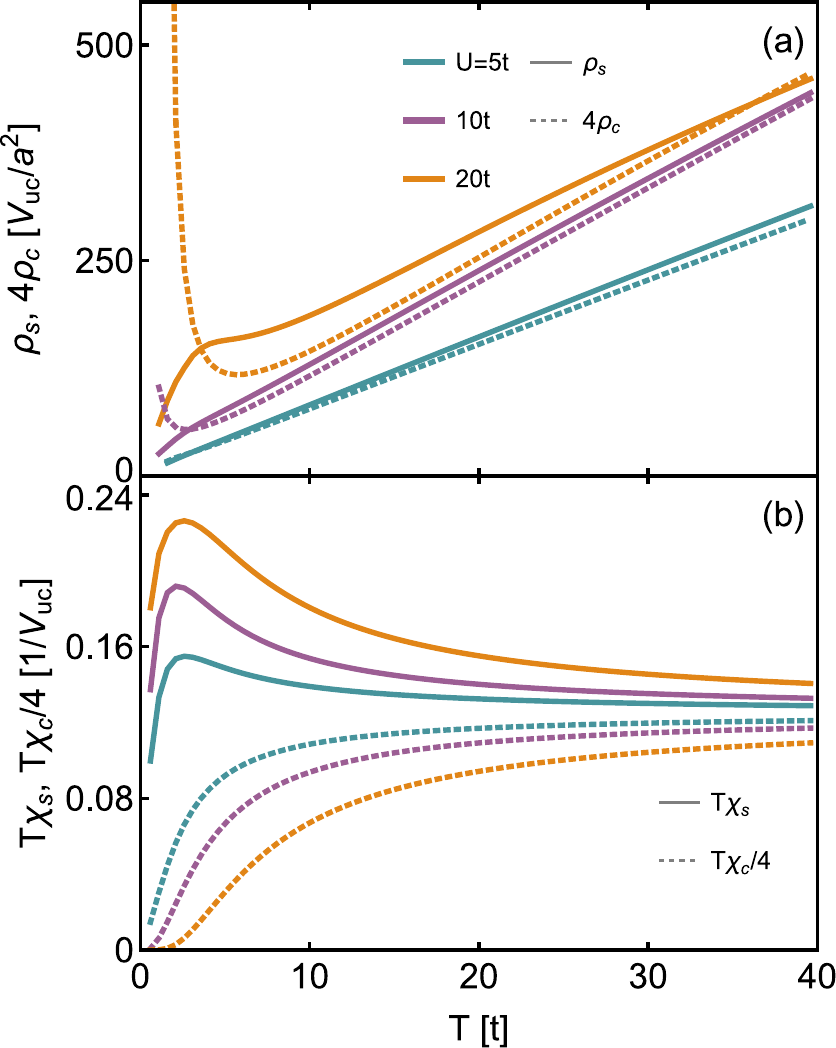}

\caption{(a): Spin and charge resistivities. The later is multiplied
  by 4 to aid the comparison. (b) Spin and charge susceptibilities
  multiplied by temperature.}
\label{fig:sup_rho}
\end{center}
\end{figure}




\section{Diffusion bound}
\label{sec-app-d}
We compare FTLM results to the conjectured lower bound on diffusion~\cite{hartnoll15}, given by
\begin{equation}
    D_s\gtrsim D_H=\frac{v_s^2}{T},
\end{equation}
\noindent
for some characteristic velocity $v_s$. While we cannot directly evaluate
the velocity, we can estimate it using the 
non-interacting-like models. We take $v_s=\frac{4Ja}{\sqrt{2}\pi}$ as a
typical velocity of spins deep in the Mott-insulating phase and
$v_s=\frac{8ta}{\sqrt{2}\pi}$ as a typical speed of electrons. The
real velocity is expected to be close to and interpolate between these
two values. Our best estimate of velocity $v_s$ is obtained via
calculation of $D_s$ and $\tau_s$ (see Fig.~\ref{fig:sup_lorentz}) and using the
approximate relation $D_s=v_s^2\tau_s/2$ as discussed in the main text.  
We use these three velocites to estimate the various diffusion bounds
and compare them with our FTLM results in  Fig.~\ref{fig:sup_hartnoll}.

\begin{figure}[ht]
\begin{center}
\includegraphics[width=0.4\textwidth]{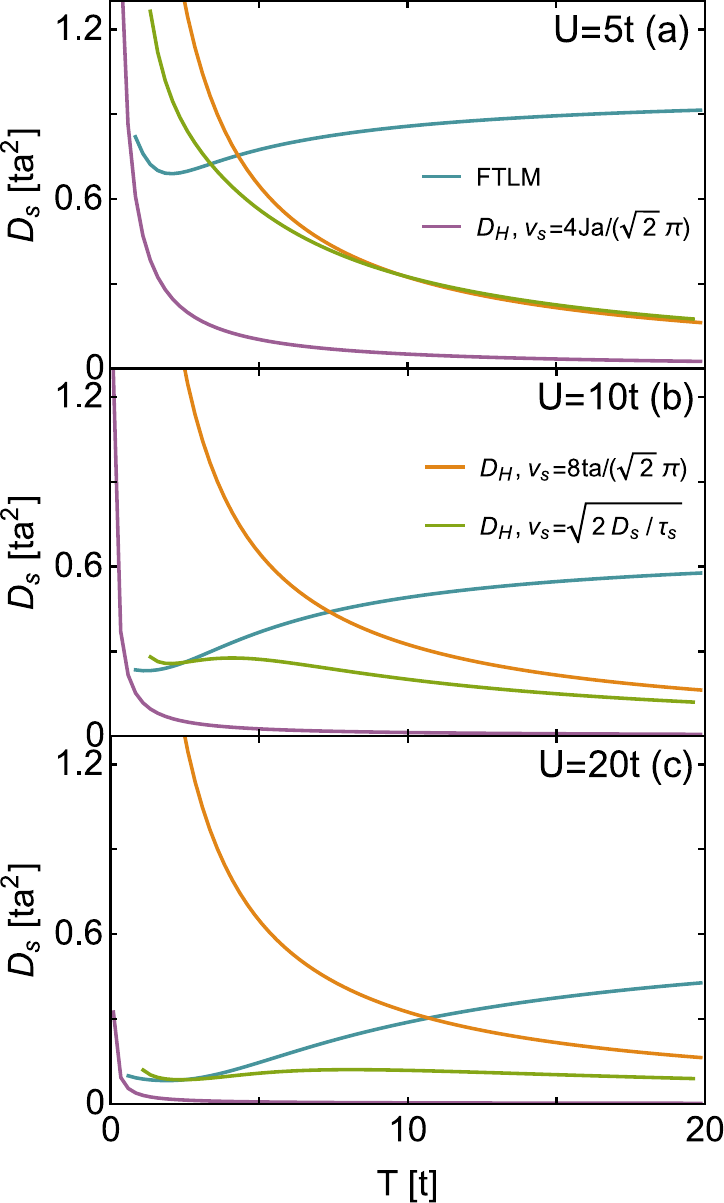}

\caption{Comparison of $D_s$ calculated with FTLM results for various
  $U$ to the conjectured diffusion bound\protect\cite{hartnoll15}
  $D_H=v_s^2/T$. If the bound is valid, the calculated $D_s$ (denoted
  with FTLM) would be
  larger than $D_H$. Due to uncertainty of $v_s$, which is needed to evaluate
  $D_H$, we compare to three values of $D_H$ obtained with three
  estimates for $v_s$. Our best estimate of $v_s$ is obtained via
  $\tau_s$ and the corresponding $D_H$ is shown with green line. At
  lowest $T$, our results indicate possible violation of the proposed
  bound.}
\label{fig:sup_hartnoll}
\end{center}
\end{figure}

We observe no apparent violation of the bound for 
$v_s=\frac{4Ja}{\sqrt{2}\pi}$,  strong violation of the bound for
$v_s=\frac{8ta}{\sqrt{2}\pi}$, while for our best estimate of
$v_s=\sqrt{2 D_s/\tau_s}$ there is an indication of diffusion bound
violation at lowest-$T$. It is most apparent for $U=5t$, as shown in
Fig.~\ref{fig:sup_hartnoll}(a). However, due to rough estimate of $v_s$,
we do not consider this as a clear-cut violation. We do, however, 
remark that the proposed bound does not manifest in any clear  way in 
our results (in distinction with the Mott-Ioffe-Regel value).

\begin{figure}[h]
\begin{center}
\includegraphics[width=0.45\textwidth]{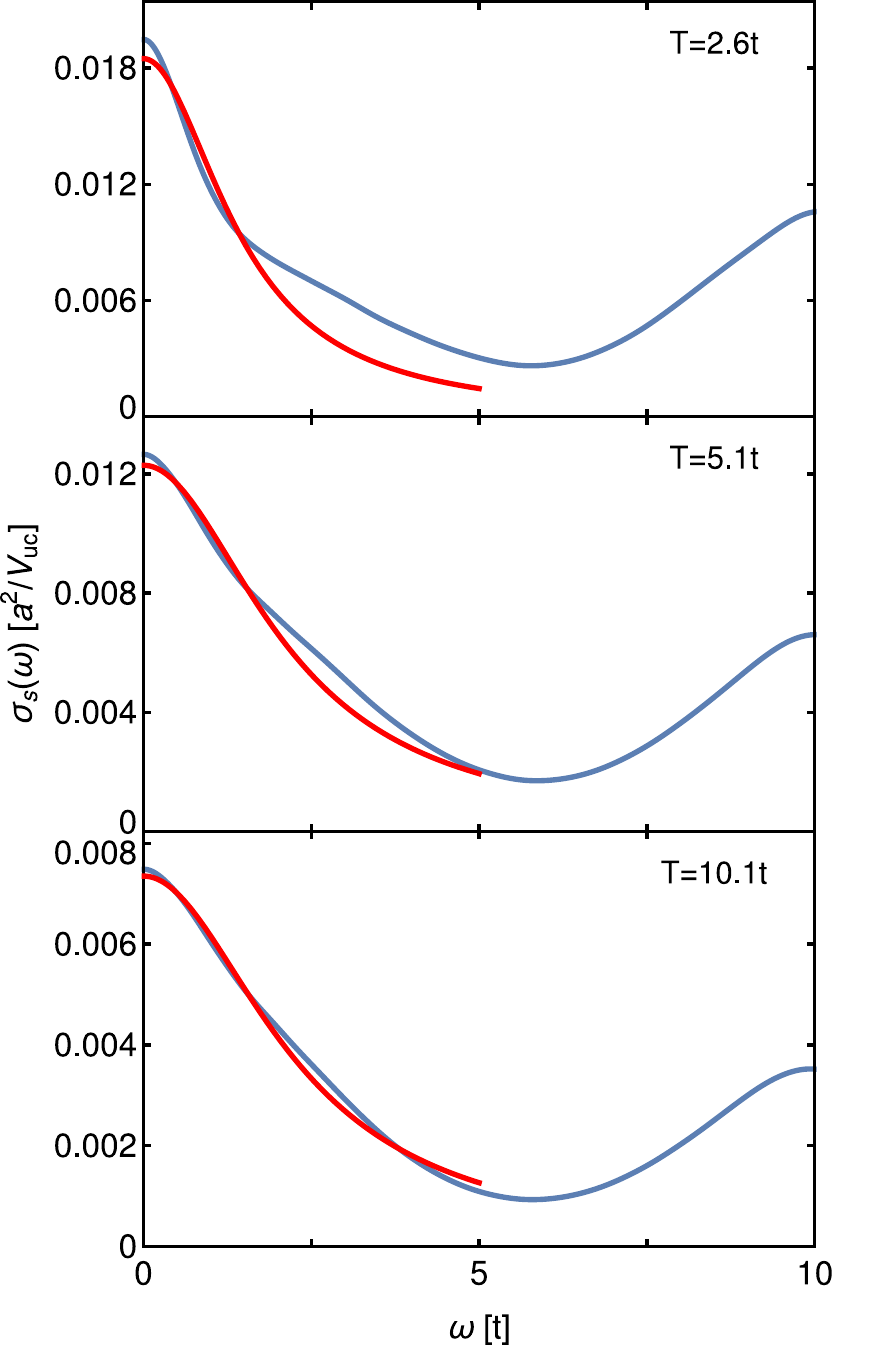}
\caption{Lorentz curve fitted to the low-$\omega$ structure
in $\sigma_s(\omega)$ for $U=10$ and several $T$. For fitting we consider only $\omega\leq 2 t$.
We estimate the uncertainty of the extracted scattering time
to be $\sim 30\%$ at low $T$ and $\sim5\%$ at $T\gtrsim U/2$.}
\label{fig:sup_lorentz}
\end{center}
\end{figure}

\section{Scattering time}
\label{sec-app-e}
To estimate the spin  scattering time, we fit the Lorentz curve to the
low-$\omega$ part of $\sigma_s(\omega)$. For illustration see
Fig.~\ref{fig:sup_lorentz}.  We choose to fit in the
frequency range from $\omega=0$ to $\omega=2t$. Due to the dependence
of extracted width ($1/\tau_s$) on frequency 
range and other possible prescriptions, e.g., taking the half width at
half maximum, our value of $\tau_s$ should be taken as a rough
estimate. We estimate its uncertainty to about 30\%, and to be largest
at lowest $T$. This uncertainty is then translated also to the
uncertainty in the spin velocity $v_s$ and mean-free-path $l_s$. 
We note that the spectra at low-$\omega$ are a bit sharper and
more linear than the Lorentz curve. Quite linear in $\omega$
behavior was, e.g., observed in the optical conductivity for the
$t$-$J$ model \cite{jaklic00}, where comparable $\omega$ resolution
can be reached with smaller number of Lanczos steps. For a recent 
discussion, see Ref. \onlinecite{schonle2020eigenstate}.

}


\begin{thebibliography}{40}%
\makeatletter
\providecommand \@ifxundefined [1]{%
 \@ifx{#1\undefined}
}%
\providecommand \@ifnum [1]{%
 \ifnum #1\expandafter \@firstoftwo
 \else \expandafter \@secondoftwo
 \fi
}%
\providecommand \@ifx [1]{%
 \ifx #1\expandafter \@firstoftwo
 \else \expandafter \@secondoftwo
 \fi
}%
\providecommand \natexlab [1]{#1}%
\providecommand \enquote  [1]{``#1''}%
\providecommand \bibnamefont  [1]{#1}%
\providecommand \bibfnamefont [1]{#1}%
\providecommand \citenamefont [1]{#1}%
\providecommand \href@noop [0]{\@secondoftwo}%
\providecommand \href [0]{\begingroup \@sanitize@url \@href}%
\providecommand \@href[1]{\@@startlink{#1}\@@href}%
\providecommand \@@href[1]{\endgroup#1\@@endlink}%
\providecommand \@sanitize@url [0]{\catcode `\\12\catcode `\$12\catcode
  `\&12\catcode `\#12\catcode `\^12\catcode `\_12\catcode `\%12\relax}%
\providecommand \@@startlink[1]{}%
\providecommand \@@endlink[0]{}%
\providecommand \url  [0]{\begingroup\@sanitize@url \@url }%
\providecommand \@url [1]{\endgroup\@href {#1}{\urlprefix }}%
\providecommand \urlprefix  [0]{URL }%
\providecommand \Eprint [0]{\href }%
\providecommand \doibase [0]{http://dx.doi.org/}%
\providecommand \selectlanguage [0]{\@gobble}%
\providecommand \bibinfo  [0]{\@secondoftwo}%
\providecommand \bibfield  [0]{\@secondoftwo}%
\providecommand \translation [1]{[#1]}%
\providecommand \BibitemOpen [0]{}%
\providecommand \bibitemStop [0]{}%
\providecommand \bibitemNoStop [0]{.\EOS\space}%
\providecommand \EOS [0]{\spacefactor3000\relax}%
\providecommand \BibitemShut  [1]{\csname bibitem#1\endcsname}%
\let\auto@bib@innerbib\@empty
\bibitem [{\citenamefont {Kokalj}(2017)}]{kokalj17}%
  \BibitemOpen
  \bibfield  {author} {\bibinfo {author} {\bibfnamefont {J.}~\bibnamefont
  {Kokalj}},\ }\href {\doibase 10.1103/PhysRevB.95.041110} {\bibfield
  {journal} {\bibinfo  {journal} {Phys. Rev. B}\ }\textbf {\bibinfo {volume}
  {95}},\ \bibinfo {pages} {041110} (\bibinfo {year} {2017})}\BibitemShut
  {NoStop}%
\bibitem [{\citenamefont {Perepelitsky}\ \emph {et~al.}(2016)\citenamefont
  {Perepelitsky}, \citenamefont {Galatas}, \citenamefont {Mravlje},
  \citenamefont {\ifmmode~\check{Z}\else \v{Z}\fi{}itko}, \citenamefont
  {Khatami}, \citenamefont {Shastry},\ and\ \citenamefont
  {Georges}}]{perepelitsky16}%
  \BibitemOpen
  \bibfield  {author} {\bibinfo {author} {\bibfnamefont {E.}~\bibnamefont
  {Perepelitsky}}, \bibinfo {author} {\bibfnamefont {A.}~\bibnamefont
  {Galatas}}, \bibinfo {author} {\bibfnamefont {J.}~\bibnamefont {Mravlje}},
  \bibinfo {author} {\bibfnamefont {R.}~\bibnamefont {\ifmmode~\check{Z}\else
  \v{Z}\fi{}itko}}, \bibinfo {author} {\bibfnamefont {E.}~\bibnamefont
  {Khatami}}, \bibinfo {author} {\bibfnamefont {B.~S.}\ \bibnamefont
  {Shastry}}, \ and\ \bibinfo {author} {\bibfnamefont {A.}~\bibnamefont
  {Georges}},\ }\href {\doibase 10.1103/PhysRevB.94.235115} {\bibfield
  {journal} {\bibinfo  {journal} {Phys. Rev. B}\ }\textbf {\bibinfo {volume}
  {94}},\ \bibinfo {pages} {235115} (\bibinfo {year} {2016})}\BibitemShut
  {NoStop}%
\bibitem [{\citenamefont {Brown}\ \emph {et~al.}(2019)\citenamefont {Brown},
  \citenamefont {Mitra}, \citenamefont {Guardado-Sanchez}, \citenamefont
  {Nourafkan}, \citenamefont {Reymbaut}, \citenamefont {H{\'e}bert},
  \citenamefont {Bergeron}, \citenamefont {Tremblay}, \citenamefont {Kokalj},
  \citenamefont {Huse}, \citenamefont {Schau{\ss}},\ and\ \citenamefont
  {Bakr}}]{brown19}%
  \BibitemOpen
  \bibfield  {author} {\bibinfo {author} {\bibfnamefont {P.~T.}\ \bibnamefont
  {Brown}}, \bibinfo {author} {\bibfnamefont {D.}~\bibnamefont {Mitra}},
  \bibinfo {author} {\bibfnamefont {E.}~\bibnamefont {Guardado-Sanchez}},
  \bibinfo {author} {\bibfnamefont {R.}~\bibnamefont {Nourafkan}}, \bibinfo
  {author} {\bibfnamefont {A.}~\bibnamefont {Reymbaut}}, \bibinfo {author}
  {\bibfnamefont {C.-D.}\ \bibnamefont {H{\'e}bert}}, \bibinfo {author}
  {\bibfnamefont {S.}~\bibnamefont {Bergeron}}, \bibinfo {author}
  {\bibfnamefont {A.-M.~S.}\ \bibnamefont {Tremblay}}, \bibinfo {author}
  {\bibfnamefont {J.}~\bibnamefont {Kokalj}}, \bibinfo {author} {\bibfnamefont
  {D.~A.}\ \bibnamefont {Huse}}, \bibinfo {author} {\bibfnamefont
  {P.}~\bibnamefont {Schau{\ss}}}, \ and\ \bibinfo {author} {\bibfnamefont
  {W.~S.}\ \bibnamefont {Bakr}},\ }\href {\doibase 10.1126/science.aat4134}
  {\bibfield  {journal} {\bibinfo  {journal} {Science}\ }\textbf {\bibinfo
  {volume} {363}},\ \bibinfo {pages} {379} (\bibinfo {year}
  {2019})}\BibitemShut {NoStop}%
\bibitem [{\citenamefont {Nichols}\ \emph {et~al.}(2019)\citenamefont
  {Nichols}, \citenamefont {Cheuk}, \citenamefont {Okan}, \citenamefont
  {Hartke}, \citenamefont {Mendez}, \citenamefont {Senthil}, \citenamefont
  {Khatami}, \citenamefont {Zhang},\ and\ \citenamefont
  {Zwierlein}}]{nichols19}%
  \BibitemOpen
  \bibfield  {author} {\bibinfo {author} {\bibfnamefont {M.~A.}\ \bibnamefont
  {Nichols}}, \bibinfo {author} {\bibfnamefont {L.~W.}\ \bibnamefont {Cheuk}},
  \bibinfo {author} {\bibfnamefont {M.}~\bibnamefont {Okan}}, \bibinfo {author}
  {\bibfnamefont {T.~R.}\ \bibnamefont {Hartke}}, \bibinfo {author}
  {\bibfnamefont {E.}~\bibnamefont {Mendez}}, \bibinfo {author} {\bibfnamefont
  {T.}~\bibnamefont {Senthil}}, \bibinfo {author} {\bibfnamefont
  {E.}~\bibnamefont {Khatami}}, \bibinfo {author} {\bibfnamefont
  {H.}~\bibnamefont {Zhang}}, \ and\ \bibinfo {author} {\bibfnamefont {M.~W.}\
  \bibnamefont {Zwierlein}},\ }\href {\doibase 10.1126/science.aat4387}
  {\bibfield  {journal} {\bibinfo  {journal} {Science}\ }\textbf {\bibinfo
  {volume} {363}},\ \bibinfo {pages} {383} (\bibinfo {year}
  {2019})}\BibitemShut {NoStop}%
\bibitem [{\citenamefont {Anderson}(1997)}]{anderson97}%
  \BibitemOpen
  \bibfield  {author} {\bibinfo {author} {\bibfnamefont {P.~W.}\ \bibnamefont
  {Anderson}},\ }\href {\doibase doi.org/10.1063/1.881959} {\bibfield
  {journal} {\bibinfo  {journal} {Phys. Today}\ }\textbf {\bibinfo {volume}
  {10}},\ \bibinfo {pages} {42} (\bibinfo {year} {1997})}\BibitemShut {NoStop}%
\bibitem [{\citenamefont {Lee}\ \emph {et~al.}(2006)\citenamefont {Lee},
  \citenamefont {Nagaosa},\ and\ \citenamefont {Wen}}]{lee06}%
  \BibitemOpen
  \bibfield  {author} {\bibinfo {author} {\bibfnamefont {P.~A.}\ \bibnamefont
  {Lee}}, \bibinfo {author} {\bibfnamefont {N.}~\bibnamefont {Nagaosa}}, \ and\
  \bibinfo {author} {\bibfnamefont {X.-G.}\ \bibnamefont {Wen}},\ }\href
  {\doibase 10.1103/RevModPhys.78.17} {\bibfield  {journal} {\bibinfo
  {journal} {Rev. Mod. Phys.}\ }\textbf {\bibinfo {volume} {78}},\ \bibinfo
  {pages} {17} (\bibinfo {year} {2006})}\BibitemShut {NoStop}%
\bibitem [{\citenamefont {\ifmmode \check{Z}\else
  \v{Z}\fi{}uti\ifmmode~\acute{c}\else \'{c}\fi{}}\ \emph
  {et~al.}(2004)\citenamefont {\ifmmode \check{Z}\else
  \v{Z}\fi{}uti\ifmmode~\acute{c}\else \'{c}\fi{}}, \citenamefont {Fabian},\
  and\ \citenamefont {Das~Sarma}}]{zutic04}%
  \BibitemOpen
  \bibfield  {author} {\bibinfo {author} {\bibfnamefont {I.}~\bibnamefont
  {\ifmmode \check{Z}\else \v{Z}\fi{}uti\ifmmode~\acute{c}\else \'{c}\fi{}}},
  \bibinfo {author} {\bibfnamefont {J.}~\bibnamefont {Fabian}}, \ and\ \bibinfo
  {author} {\bibfnamefont {S.}~\bibnamefont {Das~Sarma}},\ }\href {\doibase
  10.1103/RevModPhys.76.323} {\bibfield  {journal} {\bibinfo  {journal} {Rev.
  Mod. Phys.}\ }\textbf {\bibinfo {volume} {76}},\ \bibinfo {pages} {323}
  (\bibinfo {year} {2004})}\BibitemShut {NoStop}%
\bibitem [{\citenamefont {Chumak}\ \emph {et~al.}(2015)\citenamefont {Chumak},
  \citenamefont {Vasyuchka}, \citenamefont {Serga},\ and\ \citenamefont
  {Hillebrands}}]{chumak15}%
  \BibitemOpen
  \bibfield  {author} {\bibinfo {author} {\bibfnamefont {A.~V.}\ \bibnamefont
  {Chumak}}, \bibinfo {author} {\bibfnamefont {V.~I.}\ \bibnamefont
  {Vasyuchka}}, \bibinfo {author} {\bibfnamefont {A.~A.}\ \bibnamefont
  {Serga}}, \ and\ \bibinfo {author} {\bibfnamefont {B.}~\bibnamefont
  {Hillebrands}},\ }\href {\doibase 10.1038/nphys3347} {\bibfield  {journal}
  {\bibinfo  {journal} {Nat. Phys.}\ }\textbf {\bibinfo {volume} {11}},\
  \bibinfo {pages} {453} (\bibinfo {year} {2015})}\BibitemShut {NoStop}%
\bibitem [{\citenamefont {Hirohata}\ \emph {et~al.}(2020)\citenamefont
  {Hirohata}, \citenamefont {Yamada}, \citenamefont {Nakatani}, \citenamefont
  {Prejbeanu}, \citenamefont {Diény}, \citenamefont {Pirro},\ and\
  \citenamefont {Hillebrands}}]{hiroata20}%
  \BibitemOpen
  \bibfield  {author} {\bibinfo {author} {\bibfnamefont {A.}~\bibnamefont
  {Hirohata}}, \bibinfo {author} {\bibfnamefont {K.}~\bibnamefont {Yamada}},
  \bibinfo {author} {\bibfnamefont {Y.}~\bibnamefont {Nakatani}}, \bibinfo
  {author} {\bibfnamefont {I.-L.}\ \bibnamefont {Prejbeanu}}, \bibinfo {author}
  {\bibfnamefont {B.}~\bibnamefont {Diény}}, \bibinfo {author} {\bibfnamefont
  {P.}~\bibnamefont {Pirro}}, \ and\ \bibinfo {author} {\bibfnamefont
  {B.}~\bibnamefont {Hillebrands}},\ }\href {\doibase
  https://doi.org/10.1016/j.jmmm.2020.166711} {\bibfield  {journal} {\bibinfo
  {journal} {J. Magn. Magn. Mater.}\ }\textbf {\bibinfo {volume} {509}},\
  \bibinfo {pages} {166711} (\bibinfo {year} {2020})}\BibitemShut {NoStop}%
\bibitem [{\citenamefont {Zotos}\ \emph {et~al.}(1997)\citenamefont {Zotos},
  \citenamefont {Naef},\ and\ \citenamefont {Prelov\v{s}ek}}]{zotos97}%
  \BibitemOpen
  \bibfield  {author} {\bibinfo {author} {\bibfnamefont {X.}~\bibnamefont
  {Zotos}}, \bibinfo {author} {\bibfnamefont {F.}~\bibnamefont {Naef}}, \ and\
  \bibinfo {author} {\bibfnamefont {P.}~\bibnamefont {Prelov\v{s}ek}},\ }\href
  {\doibase 10.1103/PhysRevB.55.11029} {\bibfield  {journal} {\bibinfo
  {journal} {Phys. Rev. B}\ }\textbf {\bibinfo {volume} {55}},\ \bibinfo
  {pages} {11029} (\bibinfo {year} {1997})}\BibitemShut {NoStop}%
\bibitem [{\citenamefont {Sologubenko}\ \emph {et~al.}(2001)\citenamefont
  {Sologubenko}, \citenamefont {Giann\`o}, \citenamefont {Ott}, \citenamefont
  {Vietkine},\ and\ \citenamefont {Revcolevschi}}]{sologubenko01}%
  \BibitemOpen
  \bibfield  {author} {\bibinfo {author} {\bibfnamefont {A.~V.}\ \bibnamefont
  {Sologubenko}}, \bibinfo {author} {\bibfnamefont {K.}~\bibnamefont
  {Giann\`o}}, \bibinfo {author} {\bibfnamefont {H.~R.}\ \bibnamefont {Ott}},
  \bibinfo {author} {\bibfnamefont {A.}~\bibnamefont {Vietkine}}, \ and\
  \bibinfo {author} {\bibfnamefont {A.}~\bibnamefont {Revcolevschi}},\ }\href
  {\doibase 10.1103/PhysRevB.64.054412} {\bibfield  {journal} {\bibinfo
  {journal} {Phys. Rev. B}\ }\textbf {\bibinfo {volume} {64}},\ \bibinfo
  {pages} {054412} (\bibinfo {year} {2001})}\BibitemShut {NoStop}%
\bibitem [{\citenamefont {Hess}\ \emph {et~al.}(2001)\citenamefont {Hess},
  \citenamefont {Baumann}, \citenamefont {Ammerahl}, \citenamefont {B\"uchner},
  \citenamefont {Heidrich-Meisner}, \citenamefont {Brenig},\ and\ \citenamefont
  {Revcolevschi}}]{hess01}%
  \BibitemOpen
  \bibfield  {author} {\bibinfo {author} {\bibfnamefont {C.}~\bibnamefont
  {Hess}}, \bibinfo {author} {\bibfnamefont {C.}~\bibnamefont {Baumann}},
  \bibinfo {author} {\bibfnamefont {U.}~\bibnamefont {Ammerahl}}, \bibinfo
  {author} {\bibfnamefont {B.}~\bibnamefont {B\"uchner}}, \bibinfo {author}
  {\bibfnamefont {F.}~\bibnamefont {Heidrich-Meisner}}, \bibinfo {author}
  {\bibfnamefont {W.}~\bibnamefont {Brenig}}, \ and\ \bibinfo {author}
  {\bibfnamefont {A.}~\bibnamefont {Revcolevschi}},\ }\href {\doibase
  10.1103/PhysRevB.64.184305} {\bibfield  {journal} {\bibinfo  {journal} {Phys.
  Rev. B}\ }\textbf {\bibinfo {volume} {64}},\ \bibinfo {pages} {184305}
  (\bibinfo {year} {2001})}\BibitemShut {NoStop}%
\bibitem [{\citenamefont {Sokol}\ \emph {et~al.}(1993)\citenamefont {Sokol},
  \citenamefont {Gagliano},\ and\ \citenamefont {Bacci}}]{sokol93}%
  \BibitemOpen
  \bibfield  {author} {\bibinfo {author} {\bibfnamefont {A.}~\bibnamefont
  {Sokol}}, \bibinfo {author} {\bibfnamefont {E.}~\bibnamefont {Gagliano}}, \
  and\ \bibinfo {author} {\bibfnamefont {S.}~\bibnamefont {Bacci}},\ }\href
  {\doibase 10.1103/PhysRevB.47.14646} {\bibfield  {journal} {\bibinfo
  {journal} {Phys. Rev. B}\ }\textbf {\bibinfo {volume} {47}},\ \bibinfo
  {pages} {14646} (\bibinfo {year} {1993})}\BibitemShut {NoStop}%
\bibitem [{\citenamefont {Larionov}(2004)}]{larionov04}%
  \BibitemOpen
  \bibfield  {author} {\bibinfo {author} {\bibfnamefont {I.~A.}\ \bibnamefont
  {Larionov}},\ }\href {\doibase 10.1103/PhysRevB.69.214525} {\bibfield
  {journal} {\bibinfo  {journal} {Phys. Rev. B}\ }\textbf {\bibinfo {volume}
  {69}},\ \bibinfo {pages} {214525} (\bibinfo {year} {2004})}\BibitemShut
  {NoStop}%
\bibitem [{\citenamefont {Yusuf}\ \emph {et~al.}(2007)\citenamefont {Yusuf},
  \citenamefont {Powell},\ and\ \citenamefont {McKenzie}}]{yusuf07}%
  \BibitemOpen
  \bibfield  {author} {\bibinfo {author} {\bibfnamefont {E.}~\bibnamefont
  {Yusuf}}, \bibinfo {author} {\bibfnamefont {B.~J.}\ \bibnamefont {Powell}}, \
  and\ \bibinfo {author} {\bibfnamefont {R.~H.}\ \bibnamefont {McKenzie}},\
  }\href {\doibase 10.1103/PhysRevB.75.214515} {\bibfield  {journal} {\bibinfo
  {journal} {Phys. Rev. B}\ }\textbf {\bibinfo {volume} {75}},\ \bibinfo
  {pages} {214515} (\bibinfo {year} {2007})}\BibitemShut {NoStop}%
\bibitem [{\citenamefont {Johnson}(1993)}]{johnson93}%
  \BibitemOpen
  \bibfield  {author} {\bibinfo {author} {\bibfnamefont {M.}~\bibnamefont
  {Johnson}},\ }\href {\doibase 10.1103/PhysRevLett.70.2142} {\bibfield
  {journal} {\bibinfo  {journal} {Phys. Rev. Lett.}\ }\textbf {\bibinfo
  {volume} {70}},\ \bibinfo {pages} {2142} (\bibinfo {year}
  {1993})}\BibitemShut {NoStop}%
\bibitem [{\citenamefont {Si}(1997)}]{si97}%
  \BibitemOpen
  \bibfield  {author} {\bibinfo {author} {\bibfnamefont {Q.}~\bibnamefont
  {Si}},\ }\href {\doibase 10.1103/PhysRevLett.78.1767} {\bibfield  {journal}
  {\bibinfo  {journal} {Phys. Rev. Lett.}\ }\textbf {\bibinfo {volume} {78}},\
  \bibinfo {pages} {1767} (\bibinfo {year} {1997})}\BibitemShut {NoStop}%
\bibitem [{\citenamefont {Ji}\ \emph {et~al.}(2004)\citenamefont {Ji},
  \citenamefont {Hoffmann}, \citenamefont {Jiang},\ and\ \citenamefont
  {Bader}}]{ji04}%
  \BibitemOpen
  \bibfield  {author} {\bibinfo {author} {\bibfnamefont {Y.}~\bibnamefont
  {Ji}}, \bibinfo {author} {\bibfnamefont {A.}~\bibnamefont {Hoffmann}},
  \bibinfo {author} {\bibfnamefont {J.~S.}\ \bibnamefont {Jiang}}, \ and\
  \bibinfo {author} {\bibfnamefont {S.~D.}\ \bibnamefont {Bader}},\ }\href
  {\doibase 10.1063/1.1841455} {\bibfield  {journal} {\bibinfo  {journal} {App.
  Phys. Lett.}\ }\textbf {\bibinfo {volume} {85}},\ \bibinfo {pages} {6218}
  (\bibinfo {year} {2004})}\BibitemShut {NoStop}%
\bibitem [{\citenamefont {Meier}\ and\ \citenamefont {Loss}(2003)}]{meier03}%
  \BibitemOpen
  \bibfield  {author} {\bibinfo {author} {\bibfnamefont {F.}~\bibnamefont
  {Meier}}\ and\ \bibinfo {author} {\bibfnamefont {D.}~\bibnamefont {Loss}},\
  }\href {\doibase 10.1103/PhysRevLett.90.167204} {\bibfield  {journal}
  {\bibinfo  {journal} {Phys. Rev. Lett.}\ }\textbf {\bibinfo {volume} {90}},\
  \bibinfo {pages} {167204} (\bibinfo {year} {2003})}\BibitemShut {NoStop}%
\bibitem [{\citenamefont {Cornelissen}\ \emph {et~al.}(2015)\citenamefont
  {Cornelissen}, \citenamefont {Liu}, \citenamefont {Duine}, \citenamefont
  {Youssef},\ and\ \citenamefont {van Wees}}]{cornelissen15}%
  \BibitemOpen
  \bibfield  {author} {\bibinfo {author} {\bibfnamefont {L.~J.}\ \bibnamefont
  {Cornelissen}}, \bibinfo {author} {\bibfnamefont {J.}~\bibnamefont {Liu}},
  \bibinfo {author} {\bibfnamefont {R.~A.}\ \bibnamefont {Duine}}, \bibinfo
  {author} {\bibfnamefont {J.~B.}\ \bibnamefont {Youssef}}, \ and\ \bibinfo
  {author} {\bibfnamefont {B.~J.}\ \bibnamefont {van Wees}},\ }\href {\doibase
  10.1038/nphys3465} {\bibfield  {journal} {\bibinfo  {journal} {Nat. Phys.}\
  }\textbf {\bibinfo {volume} {11}},\ \bibinfo {pages} {1022} (\bibinfo {year}
  {2015})}\BibitemShut {NoStop}%
\bibitem [{\citenamefont {Bennett}\ and\ \citenamefont
  {Martin}(1965)}]{bennett65}%
  \BibitemOpen
  \bibfield  {author} {\bibinfo {author} {\bibfnamefont {H.~S.}\ \bibnamefont
  {Bennett}}\ and\ \bibinfo {author} {\bibfnamefont {P.~C.}\ \bibnamefont
  {Martin}},\ }\href {\doibase 10.1103/PhysRev.138.A608} {\bibfield  {journal}
  {\bibinfo  {journal} {Phys. Rev.}\ }\textbf {\bibinfo {volume} {138}},\
  \bibinfo {pages} {A608} (\bibinfo {year} {1965})}\BibitemShut {NoStop}%
\bibitem [{\citenamefont {Nagao}\ and\ \citenamefont
  {Igarashi}(1998)}]{nagao98}%
  \BibitemOpen
  \bibfield  {author} {\bibinfo {author} {\bibfnamefont {T.}~\bibnamefont
  {Nagao}}\ and\ \bibinfo {author} {\bibfnamefont {J.-i.}\ \bibnamefont
  {Igarashi}},\ }\href {\doibase 10.1143/JPSJ.67.1029} {\bibfield  {journal}
  {\bibinfo  {journal} {J. Phys. Soc. Japan}\ }\textbf {\bibinfo {volume}
  {67}},\ \bibinfo {pages} {1029} (\bibinfo {year} {1998})}\BibitemShut
  {NoStop}%
\bibitem [{\citenamefont {Sentef}\ \emph {et~al.}(2007)\citenamefont {Sentef},
  \citenamefont {Kollar},\ and\ \citenamefont {Kampf}}]{sentef07}%
  \BibitemOpen
  \bibfield  {author} {\bibinfo {author} {\bibfnamefont {M.}~\bibnamefont
  {Sentef}}, \bibinfo {author} {\bibfnamefont {M.}~\bibnamefont {Kollar}}, \
  and\ \bibinfo {author} {\bibfnamefont {A.~P.}\ \bibnamefont {Kampf}},\ }\href
  {\doibase 10.1103/PhysRevB.75.214403} {\bibfield  {journal} {\bibinfo
  {journal} {Phys. Rev. B}\ }\textbf {\bibinfo {volume} {75}},\ \bibinfo
  {pages} {214403} (\bibinfo {year} {2007})}\BibitemShut {NoStop}%
\bibitem [{\citenamefont {Pires}\ and\ \citenamefont {Lima}(2009)}]{pires09}%
  \BibitemOpen
  \bibfield  {author} {\bibinfo {author} {\bibfnamefont {A.~S.~T.}\
  \bibnamefont {Pires}}\ and\ \bibinfo {author} {\bibfnamefont {L.~S.}\
  \bibnamefont {Lima}},\ }\href {\doibase 10.1103/PhysRevB.79.064401}
  {\bibfield  {journal} {\bibinfo  {journal} {Phys. Rev. B}\ }\textbf {\bibinfo
  {volume} {79}},\ \bibinfo {pages} {064401} (\bibinfo {year}
  {2009})}\BibitemShut {NoStop}%
\bibitem [{\citenamefont {Bon\ifmmode~\check{c}\else \v{c}\fi{}a}\ and\
  \citenamefont {Jakli\ifmmode~\check{c}\else \v{c}\fi{}}(1995)}]{bonca95}%
  \BibitemOpen
  \bibfield  {author} {\bibinfo {author} {\bibfnamefont {J.}~\bibnamefont
  {Bon\ifmmode~\check{c}\else \v{c}\fi{}a}}\ and\ \bibinfo {author}
  {\bibfnamefont {J.}~\bibnamefont {Jakli\ifmmode~\check{c}\else \v{c}\fi{}}},\
  }\href {\doibase 10.1103/PhysRevB.51.16083} {\bibfield  {journal} {\bibinfo
  {journal} {Phys. Rev. B}\ }\textbf {\bibinfo {volume} {51}},\ \bibinfo
  {pages} {16083} (\bibinfo {year} {1995})}\BibitemShut {NoStop}%
\bibitem [{\citenamefont {Kopietz}(1998)}]{kopietz98}%
  \BibitemOpen
  \bibfield  {author} {\bibinfo {author} {\bibfnamefont {P.}~\bibnamefont
  {Kopietz}},\ }\href {\doibase 10.1103/PhysRevB.57.7829} {\bibfield  {journal}
  {\bibinfo  {journal} {Phys. Rev. B}\ }\textbf {\bibinfo {volume} {57}},\
  \bibinfo {pages} {7829} (\bibinfo {year} {1998})}\BibitemShut {NoStop}%
\bibitem [{\citenamefont {Jakli\v{c}}\ and\ \citenamefont
  {Prelov\v{s}ek}(2000)}]{jaklic00}%
  \BibitemOpen
  \bibfield  {author} {\bibinfo {author} {\bibfnamefont {J.}~\bibnamefont
  {Jakli\v{c}}}\ and\ \bibinfo {author} {\bibfnamefont {P.}~\bibnamefont
  {Prelov\v{s}ek}},\ }\href {\doibase 10.1080/000187300243381} {\bibfield
  {journal} {\bibinfo  {journal} {Adv. Phys.}\ }\textbf {\bibinfo {volume}
  {49}},\ \bibinfo {pages} {1} (\bibinfo {year} {2000})}\BibitemShut {NoStop}%
\bibitem [{\citenamefont {{Prelov\v{s}ek}}\ and\ \citenamefont
  {{Bon\v{c}a}}(2013)}]{prelovsek13}%
  \BibitemOpen
  \bibfield  {author} {\bibinfo {author} {\bibfnamefont {P.}~\bibnamefont
  {{Prelov\v{s}ek}}}\ and\ \bibinfo {author} {\bibfnamefont {J.}~\bibnamefont
  {{Bon\v{c}a}}},\ }\href {https://books.google.si/books?id=Be4\_AAAAQBAJ}
  {\emph {\bibinfo {title} {Strongly Correlated Systems: Numerical Methods}}},\
  edited by\ \bibinfo {editor} {\bibfnamefont {A.}~\bibnamefont {Avella}}\ and\
  \bibinfo {editor} {\bibfnamefont {F.}~\bibnamefont {Mancini}},\ Springer
  Series in Solid-State Sciences\ (\bibinfo  {publisher} {Springer Berlin
  Heidelberg},\ \bibinfo {year} {2013})\BibitemShut {NoStop}%
\bibitem [{\citenamefont {Kokalj}\ and\ \citenamefont
  {McKenzie}(2013)}]{kokalj13}%
  \BibitemOpen
  \bibfield  {author} {\bibinfo {author} {\bibfnamefont {J.}~\bibnamefont
  {Kokalj}}\ and\ \bibinfo {author} {\bibfnamefont {R.~H.}\ \bibnamefont
  {McKenzie}},\ }\href {\doibase 10.1103/PhysRevLett.110.206402} {\bibfield
  {journal} {\bibinfo  {journal} {Phys. Rev. Lett.}\ }\textbf {\bibinfo
  {volume} {110}},\ \bibinfo {pages} {206402} (\bibinfo {year}
  {2013})}\BibitemShut {NoStop}%
\bibitem [{\citenamefont {Pakhira}\ and\ \citenamefont
  {McKenzie}(2015)}]{pakhira15}%
  \BibitemOpen
  \bibfield  {author} {\bibinfo {author} {\bibfnamefont {N.}~\bibnamefont
  {Pakhira}}\ and\ \bibinfo {author} {\bibfnamefont {R.~H.}\ \bibnamefont
  {McKenzie}},\ }\href {\doibase 10.1103/PhysRevB.91.075124} {\bibfield
  {journal} {\bibinfo  {journal} {Phys. Rev. B}\ }\textbf {\bibinfo {volume}
  {91}},\ \bibinfo {pages} {075124} (\bibinfo {year} {2015})}\BibitemShut
  {NoStop}%
\bibitem [{\citenamefont {Fishman}\ and\ \citenamefont
  {Jarrell}(2002)}]{fishman02}%
  \BibitemOpen
  \bibfield  {author} {\bibinfo {author} {\bibfnamefont {R.~S.}\ \bibnamefont
  {Fishman}}\ and\ \bibinfo {author} {\bibfnamefont {M.}~\bibnamefont
  {Jarrell}},\ }\href {\doibase 10.1063/1.1456431} {\bibfield  {journal}
  {\bibinfo  {journal} {J. App. Phys.}\ }\textbf {\bibinfo {volume} {91}},\
  \bibinfo {pages} {8120} (\bibinfo {year} {2002})}\BibitemShut {NoStop}%
\bibitem [{\citenamefont {Vu\ifmmode \check{c}\else \v{c}\fi{}i\ifmmode
  \check{c}\else \v{c}\fi{}evi\ifmmode~\acute{c}\else \'{c}\fi{}}\ \emph
  {et~al.}(2019)\citenamefont {Vu\ifmmode \check{c}\else \v{c}\fi{}i\ifmmode
  \check{c}\else \v{c}\fi{}evi\ifmmode~\acute{c}\else \'{c}\fi{}},
  \citenamefont {Kokalj}, \citenamefont {\ifmmode~\check{Z}\else
  \v{Z}\fi{}itko}, \citenamefont {Wentzell}, \citenamefont
  {Tanaskovi\ifmmode~\acute{c}\else \'{c}\fi{}},\ and\ \citenamefont
  {Mravlje}}]{vucicevic19}%
  \BibitemOpen
  \bibfield  {author} {\bibinfo {author} {\bibfnamefont {J.}~\bibnamefont
  {Vu\ifmmode \check{c}\else \v{c}\fi{}i\ifmmode \check{c}\else
  \v{c}\fi{}evi\ifmmode~\acute{c}\else \'{c}\fi{}}}, \bibinfo {author}
  {\bibfnamefont {J.}~\bibnamefont {Kokalj}}, \bibinfo {author} {\bibfnamefont
  {R.}~\bibnamefont {\ifmmode~\check{Z}\else \v{Z}\fi{}itko}}, \bibinfo
  {author} {\bibfnamefont {N.}~\bibnamefont {Wentzell}}, \bibinfo {author}
  {\bibfnamefont {D.}~\bibnamefont {Tanaskovi\ifmmode~\acute{c}\else
  \'{c}\fi{}}}, \ and\ \bibinfo {author} {\bibfnamefont {J.}~\bibnamefont
  {Mravlje}},\ }\href {\doibase 10.1103/PhysRevLett.123.036601} {\bibfield
  {journal} {\bibinfo  {journal} {Phys. Rev. Lett.}\ }\textbf {\bibinfo
  {volume} {123}},\ \bibinfo {pages} {036601} (\bibinfo {year}
  {2019})}\BibitemShut {NoStop}%
\bibitem [{\citenamefont {Vrani\ifmmode~\acute{c}\else \'{c}\fi{}}\ \emph
  {et~al.}(2020)\citenamefont {Vrani\ifmmode~\acute{c}\else \'{c}\fi{}},
  \citenamefont {Vu\ifmmode \check{c}\else \v{c}\fi{}i\ifmmode \check{c}\else
  \v{c}\fi{}evi\ifmmode~\acute{c}\else \'{c}\fi{}}, \citenamefont {Kokalj},
  \citenamefont {Skolimowski}, \citenamefont {\ifmmode~\check{Z}\else
  \v{Z}\fi{}itko}, \citenamefont {Mravlje},\ and\ \citenamefont
  {Tanaskovi\ifmmode~\acute{c}\else \'{c}\fi{}}}]{vranic20}%
  \BibitemOpen
  \bibfield  {author} {\bibinfo {author} {\bibfnamefont {A.}~\bibnamefont
  {Vrani\ifmmode~\acute{c}\else \'{c}\fi{}}}, \bibinfo {author} {\bibfnamefont
  {J.}~\bibnamefont {Vu\ifmmode \check{c}\else \v{c}\fi{}i\ifmmode
  \check{c}\else \v{c}\fi{}evi\ifmmode~\acute{c}\else \'{c}\fi{}}}, \bibinfo
  {author} {\bibfnamefont {J.}~\bibnamefont {Kokalj}}, \bibinfo {author}
  {\bibfnamefont {J.}~\bibnamefont {Skolimowski}}, \bibinfo {author}
  {\bibfnamefont {R.}~\bibnamefont {\ifmmode~\check{Z}\else \v{Z}\fi{}itko}},
  \bibinfo {author} {\bibfnamefont {J.}~\bibnamefont {Mravlje}}, \ and\
  \bibinfo {author} {\bibfnamefont {D.}~\bibnamefont
  {Tanaskovi\ifmmode~\acute{c}\else \'{c}\fi{}}},\ }\href {\doibase
  10.1103/PhysRevB.102.115142} {\bibfield  {journal} {\bibinfo  {journal}
  {Phys. Rev. B}\ }\textbf {\bibinfo {volume} {102}},\ \bibinfo {pages}
  {115142} (\bibinfo {year} {2020})}\BibitemShut {NoStop}%
\bibitem [{\citenamefont {Chakravarty}\ \emph {et~al.}(1989)\citenamefont
  {Chakravarty}, \citenamefont {Halperin},\ and\ \citenamefont
  {Nelson}}]{chakravarty89}%
  \BibitemOpen
  \bibfield  {author} {\bibinfo {author} {\bibfnamefont {S.}~\bibnamefont
  {Chakravarty}}, \bibinfo {author} {\bibfnamefont {B.~I.}\ \bibnamefont
  {Halperin}}, \ and\ \bibinfo {author} {\bibfnamefont {D.~R.}\ \bibnamefont
  {Nelson}},\ }\href {\doibase 10.1103/PhysRevB.39.2344} {\bibfield  {journal}
  {\bibinfo  {journal} {Phys. Rev. B}\ }\textbf {\bibinfo {volume} {39}},\
  \bibinfo {pages} {2344} (\bibinfo {year} {1989})}\BibitemShut {NoStop}%
\bibitem [{\citenamefont {Kim}\ and\ \citenamefont {Troyer}(1998)}]{kim98}%
  \BibitemOpen
  \bibfield  {author} {\bibinfo {author} {\bibfnamefont {J.-K.}\ \bibnamefont
  {Kim}}\ and\ \bibinfo {author} {\bibfnamefont {M.}~\bibnamefont {Troyer}},\
  }\href@noop {} {\bibfield  {journal} {\bibinfo  {journal} {Phys. Rev. Lett.}\
  }\textbf {\bibinfo {volume} {80}},\ \bibinfo {pages} {2705} (\bibinfo {year}
  {1998})}\BibitemShut {NoStop}%
\bibitem [{\citenamefont {Prelov\ifmmode~\check{s}\else \v{s}\fi{}ek}\ \emph
  {et~al.}(2015)\citenamefont {Prelov\ifmmode~\check{s}\else \v{s}\fi{}ek},
  \citenamefont {Kokalj}, \citenamefont {Lenar\ifmmode \check{c}\else
  \v{c}\fi{}i\ifmmode~\check{c}\else \v{c}\fi{}},\ and\ \citenamefont
  {McKenzie}}]{prelovsek15}%
  \BibitemOpen
  \bibfield  {author} {\bibinfo {author} {\bibfnamefont {P.}~\bibnamefont
  {Prelov\ifmmode~\check{s}\else \v{s}\fi{}ek}}, \bibinfo {author}
  {\bibfnamefont {J.}~\bibnamefont {Kokalj}}, \bibinfo {author} {\bibfnamefont
  {Z.}~\bibnamefont {Lenar\ifmmode \check{c}\else
  \v{c}\fi{}i\ifmmode~\check{c}\else \v{c}\fi{}}}, \ and\ \bibinfo {author}
  {\bibfnamefont {R.~H.}\ \bibnamefont {McKenzie}},\ }\href {\doibase
  10.1103/PhysRevB.92.235155} {\bibfield  {journal} {\bibinfo  {journal} {Phys.
  Rev. B}\ }\textbf {\bibinfo {volume} {92}},\ \bibinfo {pages} {235155}
  (\bibinfo {year} {2015})}\BibitemShut {NoStop}%
\bibitem [{\citenamefont {Hartnoll}(2015)}]{hartnoll15}%
  \BibitemOpen
  \bibfield  {author} {\bibinfo {author} {\bibfnamefont {S.~A.}\ \bibnamefont
  {Hartnoll}},\ }\href {\doibase 10.1038/nphys3174} {\bibfield  {journal}
  {\bibinfo  {journal} {Nat. Phys.}\ }\textbf {\bibinfo {volume} {11}},\
  \bibinfo {pages} {54} (\bibinfo {year} {2015})}\BibitemShut {NoStop}%
\bibitem [{\citenamefont {Takigawa}\ \emph {et~al.}(1996)\citenamefont
  {Takigawa}, \citenamefont {Asano}, \citenamefont {Ajiro}, \citenamefont
  {Mekata},\ and\ \citenamefont {Uemura}}]{takigawa96}%
  \BibitemOpen
  \bibfield  {author} {\bibinfo {author} {\bibfnamefont {M.}~\bibnamefont
  {Takigawa}}, \bibinfo {author} {\bibfnamefont {T.}~\bibnamefont {Asano}},
  \bibinfo {author} {\bibfnamefont {Y.}~\bibnamefont {Ajiro}}, \bibinfo
  {author} {\bibfnamefont {M.}~\bibnamefont {Mekata}}, \ and\ \bibinfo {author}
  {\bibfnamefont {Y.~J.}\ \bibnamefont {Uemura}},\ }\href {\doibase
  10.1103/PhysRevLett.76.2173} {\bibfield  {journal} {\bibinfo  {journal}
  {Phys. Rev. Lett.}\ }\textbf {\bibinfo {volume} {76}},\ \bibinfo {pages}
  {2173} (\bibinfo {year} {1996})}\BibitemShut {NoStop}%
\bibitem [{\citenamefont {Shastry}(2008)}]{shastry09}%
  \BibitemOpen
  \bibfield  {author} {\bibinfo {author} {\bibfnamefont {B.~S.}\ \bibnamefont
  {Shastry}},\ }\href {\doibase 10.1088/0034-4885/72/1/016501} {\bibfield
  {journal} {\bibinfo  {journal} {Reports on Progress in Physics}\ }\textbf
  {\bibinfo {volume} {72}},\ \bibinfo {pages} {016501} (\bibinfo {year}
  {2008})}\BibitemShut {NoStop}%
\bibitem [{\citenamefont {Sch{\"o}nle}\ \emph {et~al.}(2020)\citenamefont
  {Sch{\"o}nle}, \citenamefont {Jansen}, \citenamefont {Heidrich-Meisner},\
  and\ \citenamefont {Vidmar}}]{schonle2020eigenstate}%
  \BibitemOpen
  \bibfield  {author} {\bibinfo {author} {\bibfnamefont {C.}~\bibnamefont
  {Sch{\"o}nle}}, \bibinfo {author} {\bibfnamefont {D.}~\bibnamefont {Jansen}},
  \bibinfo {author} {\bibfnamefont {F.}~\bibnamefont {Heidrich-Meisner}}, \
  and\ \bibinfo {author} {\bibfnamefont {L.}~\bibnamefont {Vidmar}},\
  }\href@noop {} {\bibfield  {journal} {\bibinfo  {journal} {arXiv preprint
  arXiv:2011.13958}\ } (\bibinfo {year} {2020})}\BibitemShut {NoStop}%
\end{thebibliography}

%

\end{document}